\documentclass[lettersize,journal]{IEEEtran}
\IEEEoverridecommandlockouts
\usepackage{epsfig,rotating,setspace,latexsym,amsmath,epsf,amssymb,amsfonts,bm,theorem,subfigure,epstopdf,cite,authblk,nicefrac,caption,bbm,algcompatible,algorithm,color,mathtools,multirow,soul,cancel}

\algrenewcommand\algorithmicforall{\textbf{foreach}}
\algrenewcommand\algorithmicindent{.8em}

\newtheorem{theorem}{Theorem}

\newtheorem{problem}{Problem}
\newtheorem{remark}{Remark}

\IEEEoverridecommandlockouts
\allowdisplaybreaks

\definecolor{green1}{rgb}{0.2,0.7,0.5}

\newcommand{\opS}{\operatorname{S}}

\newcommand{\opq}{\operatorname{q}}
\newcommand{\opL}{\operatorname{L}}
\newcommand{\opu}{\operatorname{u}}

\begin{document}

\title{Distributed Offloading in Multi-Access Edge Computing Systems: A Mean-Field Perspective}

\author{Shubham Aggarwal, Muhammad Aneeq uz Zaman, Melih Bastopcu, Sennur Ulukus, and Tamer~Ba{\c s}ar
\thanks{Research of SA, MAZ, MB, and TB was supported in part by the ARO MURI Grant AG285 and in part by the AFOSR Grant FA9550-24-1-0152. Research of MB was partially supported by BİL2: BILKENT University-TUBITAK BILGEM Consultancy Call for Research EDGE-4-IoT and TUBITAK 2232-B International Fellowship for Early Stage Researchers.

Shubham Aggarwal is with the Coordinated Science Laboratory and the Department of Mechanical Science and Engineering at the University of Illinois at Urbana-Champaign (UIUC); Muhammad Aneeq uz Zaman is with Analog Devices Inc.; Melih Bastopcu is with the Department of Electrical and Electronics Engineering at Bilkent University;  Sennur Ulukus is with the Department of Electrical and Computer Engineering at University of Maryland, College Park; Tamer Ba{\c s}ar is with the Coordinated Science Laboratory and the Department of Electrical and Computer Engineering at UIUC. Emails:\texttt{\{sa57, mazaman2, basar1\}@illinois.edu), \texttt{bastopcu@bilkent.edu.tr}, \texttt{ulukus@umd.edu}.}}
\thanks{An earlier, preliminary version \cite{aggarwal2024mean} of this work was presented at the 2024 IEEE Globecom Conference Workshop on Emerging Topics in 6G Communications,  December 8-12, 2024, Cape Town, South Africa.}}

\maketitle

\begin{abstract}
With the widespread adoption of internet-of-things (IoT) devices capable of supporting numerous intelligent applications, the demand for computational power has surged dramatically. Multi-access edge computing (MEC) technology is a promising solution to assist the often power-constrained IoT devices by providing additional computing resources for time-sensitive tasks. In this paper, we consider  the problem of optimal task offloading in MEC systems with due consideration of the timeliness and scalability issues under two scenarios of equitable and priority access to the edge server (ES). In the first scenario, we consider a MEC system consisting of $N$ devices assisted by one ES, where the devices can split task execution between a local processor and the ES, with \textit{equitable access} to the ES. In the second scenario, we consider a MEC system consisting of one primary user, $N$ secondary users and one ES. The primary user has \textit{priority access} to the ES while the secondary users have \textit{equitable access} to the ES amongst themselves. In both scenarios, due to the power consumption associated with utilizing the local resource and task offloading, the devices must optimize their actions. Additionally, since the ES is a shared resource, other users' offloading activity serves to increase latency incurred by each user. We thus model both scenarios using a \textit{large user} non-cooperative game framework. However, the presence of a large number of users makes it nearly impossible to compute the equilibrium offloading policies for each user, which would require a significant communication overhead to exchange information with each other. Thus, to alleviate such scalability issues, we invoke the paradigm of mean-field games to design completely distributed low complexity algorithms for the computation of approximate Nash equilibrium policies for each user based on only their local information. Further, by leveraging the novel age of information (AoI) metric, we study the trade-offs between increasing information freshness and reducing power consumption for each user. Using numerical evaluations, we show that our approach can recover the offloading trends displayed under centralized solutions, and provide additional insights into the results obtained.
\end{abstract}

\section{Introduction}
The multi-access edge computing (MEC) technology has recently garnered significant attention as a potential solution to enhance computing capabilities in power-limited internet-of-things (IoT) devices\cite{mach2017mobile}. The MEC architecture capitalizes on the benefits of wireless communication and mobile computing paradigms, thereby enabling the offloading of task execution to the network edge. This is in contrast with the traditional cloud computing technology, which is (1) located geographically far from the end-users, and (2) have limited number of computing platforms. The MEC technology, on the other hand, brings computing capability to the edge of the network, which is much closer to the consumer, increasing the number of computing stations, each serving a smaller number of users within a certain spatial area. This approach is expected to be crucial in time-sensitive applications such as vehicle positioning in autonomous driving, task assignment in warehouses, and remote surgery systems \cite{mach2017mobile, hua2018energy, muhammad2021minimizing}.

\begin{figure}[t]
	\centerline{\includegraphics[width=0.9\columnwidth]{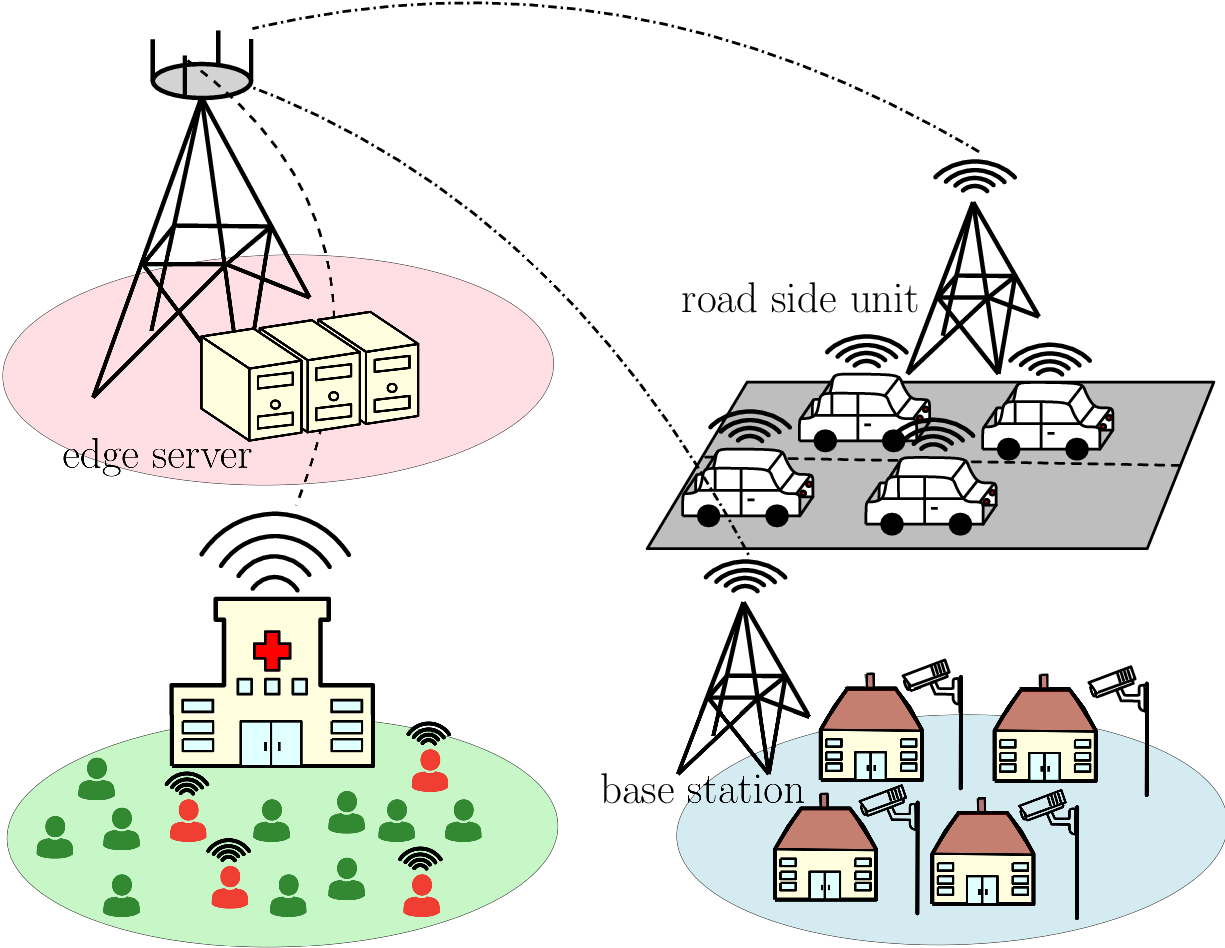}}
	\caption{\small{The figure shows a prototypical MEC system consisting of an edge server and intelligent applications such as connected autonomy, medical internet-of-things and surveillance that simultaneously utilize edge server for timely computation.} }
    \label{Fig:system_model}
    \vspace{-0.3cm}
\end{figure}

In this work, we aim to (1) increase the time responsiveness of task execution in MEC systems by employing the novel age of information (AoI) metric \cite{kaul2012real} to ensure the timeliness of task execution in time-sensitive applications, and (2) promote scalability of the paradigm by constructing completely distributed policy design methods using mean-field games (MFGs). Specifically, we first consider a MEC system with $N$ devices assisted by an edge server (ES), which can be equitably accessed by each device.\footnote{We use the words user and device interchangeably in the sequel.} Such a prototypical MEC system is shown in Fig.~\ref{Fig:system_model}. The devices aim to optimally utilize the onboard processor power and the computation facility provided by the ES. Since the ES is a shared resource, the offloading policy of each user is affected by those of the other users. Thus, we aim to find Nash equilibrium policies for each user to balance between the power consumed at each device and the timeliness incurred by it. To capture the latter, we invoke the novel AoI metric to set up each device's problem as a multi-objective optimization problem of balancing power consumption and the average AoI incurred by the offloaded packets. 
Additionally, to alleviate the issue of computational tractability of equilibrium policy computation posed by a large number of devices, we leverage the framework of MFGs \cite{huang2004uplink, huang2006large, huang2007large, lasry2007mean, wang2014mean, aggarwal2023weighted, aggarwal2023large} to compute distributed equilibrium policies based on each user's local information only. This allows us to decrease the communication overhead incurred as a result of information exchange required between users to obtain a Nash equilibrium solution. Consequently, our algorithm can scale to a large user population without additional costs. In addition, motivated by scenarios in \textit{cognitive radio network} technology \cite{priyadarshi2024techniques}, we also extend the aforementioned MEC architecture (with equitable access) to the one constituting a primary user, who has priority access to the ES while all other $N$ users have equitable (but secondary) access to the ES via the primary user's transmitter. We model this problem, again, within a non-cooperative game framework, and consequently invoke the paradigm of major-minor mean-field games (MM-MFGs)\cite{nourian2013, firoozi2017execution} to alleviate the computational tractability issue posed by the large number of users. We provide distributed algorithms to compute (local) Nash equilibrium policies for the cases with and without the primary user. Finally, we corroborate the theoretical findings using extensive numerical evaluations. 

\subsection{Related Works} 
The subject of computation offloading in MEC systems has received wide attention in the past decade or so. One line of research in the area has focused on dedicated resource allocation of a portion of the total bandwidth to the involved users \cite{mao2016power,wang2017joint,maharjan2019}. Specifically, \cite{mao2016power} formulated the offloading problem as a power consumption minimization problem with constraints on the task buffer stability; with extensions provided in \cite{jia2022lyapunov,maharjan2019} to incorporate wireless power transfer to the IoT devices. Consequently, the authors provided online algorithms that determine the local execution and computation offloading policy based on the Lyapunov penalty-plus-drift-based optimization technique \cite{neely2022stochastic} and study the trade-offs between the power consumed and delay incurred, as a function of a control parameter. While the above works do not take into account timeliness considerations serving time-critical applications, a few recent works \cite{kuang2019age, liu2021optimizing, muhammad2021minimizing} have focused on timeliness within the realm of the MEC resource allocation. Specifically, \cite{kuang2019age} considered a single-source single-destination MEC system for timely status updating; \cite{liu2021optimizing} leveraged energy harvesting in addition to the MEC to support computing capabilities of the IoT devices; and \cite{muhammad2021minimizing} jointly assessed the impact of stochastic arrivals, scheduling policy, and unreliable channel conditions on the expected sum of AoI in MEC assisted IoT networks. We further refer the interested readers to surveys \cite{lin2019computation,feng2022computation} for additional details.

In contrast, to solve the high time-complexity issues faced by the central resource allocation techniques \cite{zhou2020partial}, relatively recent works use the game-theoretic framework \cite{ning2020mobile,pham2022partial,teng2022game,zhou2022stackelberg} for designing Nash or Stackelberg equilibrium offloading policies, which take into account the self-interested nature of the involved users. Specifically, the work \cite{ning2020mobile} develops a cooperative-competitive game based optimization problem to compute offloading policies for monitoring health in internet-of-medical-things applications; \cite{pham2022partial} considers offloading decisions to minimize the delay subject to an offloading cost and design incentive schemes to utilize parked vehicles as edge computers within the MEC system; \cite{teng2022game} considers a non-sequential offloading strategy based on a non-cooperative game approach which is effective in reducing latency compared to the traditional way of transmitting, planning, forwarding, and executing sequentially; and \cite{zhou2022stackelberg} formulates an offloading problem constituting a cloud server and multiple edge servers and designs incentive mechanisms for utility maximization of all the servers in a Stackelberg game setting.

Computation of Nash equilibria in the above settings, however, poses a challenge where each user is required to have the knowledge of the policy structure of the other users, since their cost functions are coupled by the presence of a shared resource. This requires not only an additional storage mechanism, but also a significant communication overhead for information exchange, especially for a large user population. This makes it difficult to scale the MEC paradigm to large user settings---a feature which is core to the MEC paradigm. To alleviate this issue of tractable policy computation, the framework of MFGs was proposed in the control systems literature, independently and concurrently in the works \cite{huang2006large,huang2007large} and \cite{lasry2007mean} which, under suitable assumptions of homogeneity and anonymity among users, allows the computation of approximate distributed Nash equilibrium solutions without the necessity of information exchange between users. Using motivations from statistical physics, it exploits a limiting system (called a mean-field system) with $N = \infty$ to compute consistent equilibrium solutions for a ``generic'' user who is representative of the entire population. Such a framework has been widely applied in various large-user domains such as epidemiology, power systems, semantic control systems, wireless networks, satellite communications,  finance \cite{aggarwal2023weighted,aggarwal2023large,Kang2024, olmez2022modeling,carmona2020applications,al2015joint}, to name a few. Additionally, the setup of MFGs has also been extended to cases constituting the presence of an ``influential'' player in addition to the large population of ``minor'' players and is termed as major-minor MFGs \cite{nourian2013,firoozi2017execution}. Inspired by financial markets and banking systems, such a formulation allows the study of the non-vanishing effects of players which hold major portfolios within the population.

Within the context of MEC systems, to the best of the authors' knowledge, the only work employing the \textit{mean-field type game} paradigm is \cite{banez2020mean} which is comprised of end-users offloading tasks, an intermediate task aggregator which pre-processes and stores all the tasks, and multiple ESs which can ``pull'' tasks for computation. Thus, decision making is carried out from the perspective of the ESs which decide on the amount of tasks that \textit{they} can complete based on their energy and resource constraints. In contrast, in our current work, decision making is carried out at the end-user level, which decides on how much to execute locally and how much to offload, and without the presence of any intermediate authority like the task aggregator. Furthermore, our objective is based on the AoI based performance metric as opposed to \cite{banez2020mean} considering a penalty \textit{for the edge node} based on how much edge resource is utilized. Thus, our work is the first to exploit the MFG and MM-MFG framework to serve time-critical MEC systems with and without priority access. 


The main contributions of our work are as follows:
\begin{enumerate}
    \item First, we provide a novel formulation of the computation offloading problem in an (\textit{equitable access}) MEC system using AoI objective within the paradigm of non-cooperative game theory, to appropriately take into account the selfish nature of the end-users and to better support time-critical applications in comparison to traditional performance metrics of delay and throughput (see for instance, \cite{muhammad2021minimizing,kaul2012real}). The novelty further lies in that our formulation allows for direct optimization of the AoI-based objective by being able to compute closed-form expressions for the average AoI. 

    \item To alleviate the associated issues of scalability and tractable Nash equilibrium computation within a large-user system, we employ the novel MFGs framework to provide low complexity algorithms to compute \textit{completely distributed approximate Nash equilibrium} offloading policies for the energy-constrained users. This plays a significant part in reducing a) the storage requirement, and b) the communication overhead, due to information exchange between the large number of users.

    \item We also extend the framework in 1) to a novel setting of MEC with priority access motivated by techniques in cognitive radio networks, where end-users are labeled as primary and secondary based on the priority of access given to them. For the former problem, we borrow the elegant framework of major-minor MFGs from the control theory literature to, again, provide low-complexity algorithms to compute distributed Nash equilibrium offloading policies for both the primary and the secondary users.

    \item We finally present an extensive numerical analysis to validate our theoretical results and interestingly demonstrate that our formulation results in the well-studied $[O(V),O(1/V)]$ power-execution delay trade-off. Further, our MFG-based approach closely approximates Nash equilibrium policies with a significant advantage of being fully decentralized. Finally, we also demonstrate that with our primary-secondary setup, we can greatly improve the effective resource utilization by 37\%.
\end{enumerate}

Distinctively, our approach provides a very clean \textit{recipe} on how to design offloading policies for MEC systems rather than heuristic-based ones. This further goes beyond the service disciplines considered in this work such as priority-based disciplines and queueing-based ones, among others. In the process, we also provide complimentary results on the computation of average AoIs for systems with series-sum-parallel connections of servers. These extend the results of the works \cite{yates2018status,kaul2020timely} and would be of independent interest.

\textbf{Notations:} $[N]:=\{1,\ldots,N\}$ denotes the set of agents. We use the shorthand $\text{Poi}(a)$ and $\text{exp}(a)$ to denote Poisson and exponential random variables, respectively. For a policy vector $a = [a_1, \cdots, a_N]$, $a_{-i}$ denotes the policy vector of all users other than $i$.

\section{MEC with Equitable Access--Problem Setup}\label{sec:MEC_formulation}

\begin{figure}[t]
    \centering
    \includegraphics[width=0.9\columnwidth]{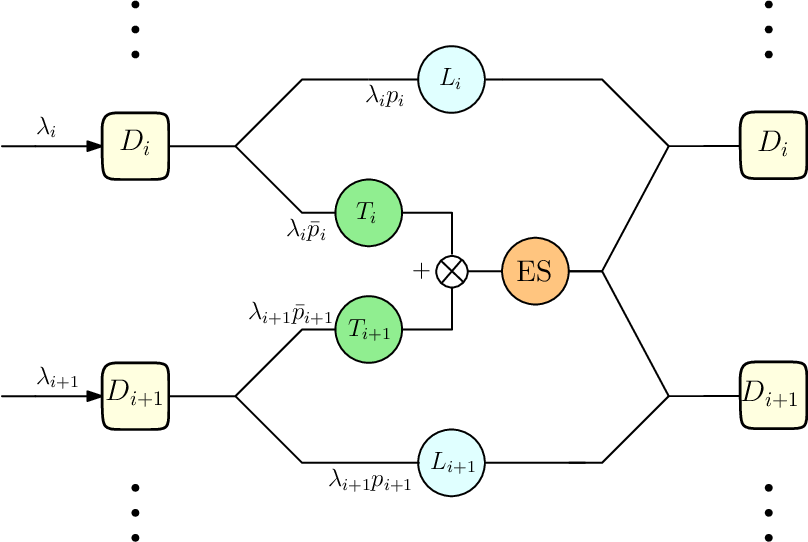}
    \caption{\small{The flow of incoming tasks is shown for a system of $N$ devices. In particular, for device $D_i$: $L_i$ and $T_i$ denote the device's local processor and its transmitter, resp.; ES denotes the edge server; $p_i$ and $\bar p_i \!=\! 1-p_i$ denote the Bernoulli probability according to which a task is either chosen to be served locally or offloaded to the ES.}}
    \label{fig:N_user}
    \vspace{-0.2cm}
\end{figure}

In this section, we start by formulating the MEC problem where each device has equitable (one-hop) access to the ES. A schematic of the system model is shown in Fig. \ref{fig:N_user} which comprises $N$ users and one ES. Associated with each user is a type variable $\phi$, which belongs to a finite set $\Phi$, and is sampled according to a probability distribution $\mathbb{P}_N(\phi)$ with $\lim_{N \rightarrow \infty} \mathbb P_N(\phi) = \mathbb P(\phi), \forall \phi$. The notion of \textit{type} of a user allows us to introduce heterogeneity within the MEC system where each device belonging to a different type can have possibly different rates of arrival, service rates, and other device characteristics. 

Next, each device $D_i$ needs to execute tasks arriving to the device. To assist in the execution process, an ES is available. Thus, each device has two task processing options: (1) it can run its own local processor $L_i$ at a certain frequency, or (2) it can use its transmitter $T_i$ to offload the computation to the ES, as shown in Fig. \ref{fig:N_user}. The inter-arrival times of tasks arriving at the $i$th device $D_i$ are distributed as an $\text{exp}(\lambda_i)$ random variable (r.v.) for all $i \in [N]$. If $D_i$ decides to carry out the tasks on $L_i$, then it can operate the processor at a frequency $\mu_{1i} \leq f_{i,max}$. The service time of $L_i$ is distributed as an $\text{exp}(\mu_{1i})$ r.v. Accordingly, the \emph{processing power} used is $P_{\ell,i} = \eta \mu_{1i}^3$, where $\eta_i$ is a positive constant denoting the processor's effective capacitance \cite{mao2016power}.

On the other hand, if $D_i$ decides to offload the task to the ES, then it gets served sequentially by  $T_i$ to the ES and the ES uploads it back to the device after processing. The transmission rate of $T_i$ is modeled as an $\text{exp}(\mu_{2i})$ r.v. with $\mu_{2i}$ being the mean \emph{transmission power usage} and $\mu_{2i} \leq P_{i,max}$. The service time of task processing at the ES is modeled as an $\text{exp}(\mu_3)$ r.v. where $\mu_3 \gg \mu_{1i}, \mu_{2i}$. We employ the \textit{last-come-first-serve with preemption} (LCFS-P) discipline\footnote{The motivation behind using a preemption based discipline is two-fold: 1) it allows for \textit{efficient} operation of systems with shared resources and selfish users (quantified using the price of anarchy or the price of stability metrics) as observed in the literature \cite{gai2016packet}, and 2) it allows for a manageable state space to compute average AoI expressions for a system with a hybrid connection of series-parallel servers.
} 
at all the servers ($L_i,T_i$, and the ES) and assume that the downloading time of the processed task by the device is negligible.\footnote{Motivated by autonomous vehicular systems or real-time monitoring systems, the uploaded tasks usually consist of high-quality images or videos which take non-negligible transmission duration versus the processed tasks, which constitute low-size commands (such as ``accident ahead'' signal, or the target's real-time position) which can be transmitted back to the IoT devices instantaneously. Also, the ES can be directly connected to the power source, and thus, it can use significantly higher transmission power. However, since IoT devices have finite batteries, their transmission times may not be negligible.}

Since the effective service rates provided by $L_i$ and the series path of $T_i$ and the ES are heterogeneous, we employ the i.i.d.~Bernoulli distributed random variable with a mean $p_i$ to split the incoming Poisson process into two independent Poisson processes with respective means $\lambda_i p_i$ and $\lambda_i \Bar{p}_i$, where $\Bar{p}_i  =1-p_i$ (as in Fig.~\ref{fig:N_user}). Such a Bernoulli splitting process has been widely employed in the literature in systems with heterogeneous parallel connection of servers \cite{yates2018status}. Finally, we measure the freshness of processed information at the device using the average AoI metric which has been widely employed in the literature \cite{kaul2012real, Bastopcu_distortion, Buyukates_dist_comp, Li_MEC_UAV, Song_MEC, sathyavageeswaran2024timely, yates2020age, sun2022age}. Formally, the AoI at the receiver (which is the device itself in our case) is  defined as the time elapsed at the receiving end since the latest delivered information packet was generated at the source.

Thus, the aim of each device $D_i$ is to find optimal offloading policies (i.e., choosing the decision variables $p_i,\mu_{1i},\mu_{2i}$) to serve a two-fold objective of: (1) minimizing the average AoI of the tasks, and (2) minimizing the power consumed during local processing and transmission. Since this is a multi-objective optimization problem, in the sequel we use the scalarization approach \cite{boyd2004convex} to set up each device's problem. Let us define $\bm{\mu}_1:= [\mu_{11}, \cdots, \mu_{1N}]$, $\bm{\mu}_2:= [\mu_{21}, \cdots, \mu_{2N}]$ and $\bm{p}:= [p_1, \cdots, p_N]$. Then, the fraction of time that $L_i$ is busy can be computed as $t_{L_i} = \nicefrac{\lambda_i p_i}{(\lambda_i p_i + \mu_{2i})}$ and the fraction of time that $T_i$ is busy can be computed as $t_{T_i} = \nicefrac{\lambda_i \bar p_i}{(\lambda_i \bar p_i + \mu_{1i})}$. Consequently, each device $i \in [N]$ wishes to solve the following problem.

\begin{problem}[$N$-user game problem]\label{problem:N_user_game}
Each device $i \in [N]$ aims to minimize its cost $J^{N,i}$:
    \begin{align*}
      & \min_{(p_i,\mu_{1i},\mu_{2i}) \in [0,1] \times \mathbb{R}^2} J^{N,i}(\boldsymbol{p},\boldsymbol{\mu_1},\boldsymbol{\mu_2}) \\
        & \qquad ~~\mbox{s.t.} ~~~0\leq \mu_{1i} \leq P_{i,max} \\
        & \qquad ~~~~~~~~ 0\leq \mu_{2i} \leq f_{i,max},
\end{align*}
where 
\begin{align*}
J^{N,i}(\boldsymbol{p},\boldsymbol{\mu_1},\boldsymbol{\mu_2}) :=  t_{T_i}\mu_{1i} +  t_{L_i} \eta \mu_{2i}^3 + V_i \Delta^{(N)}_{i}(\boldsymbol{p}, \boldsymbol{\mu_1},\boldsymbol{\mu_2}). 
\end{align*}
\end{problem}

Here, the first two terms in $J^{N,i}(\boldsymbol{p},\boldsymbol{\mu_1},\boldsymbol{\mu_2})$ denote the average power consumed at both $T_i$ and $L_i$, respectively, and $\Delta^{(N)}_{i}(\cdot, \cdot, \cdot)$ denotes the average AoI incurred by $D_i$ and $V_i>0$ is the scalarization parameter
which weights information freshness versus power consumption. A high value of $V_i$ indicates that the device $i$ cares about time responsiveness more than the power consumed and vice versa. 

We note that the Problem \ref{problem:N_user_game} is a game problem due to the presence of other devices' policies in the cost optimization problem of the $i$th device. This requires each device to know the policies of the other devices to compute its own, which can incur a significant communication overhead, especially in a large-user scenario. Thus, we will later employ the MFG framework to alleviate this issue and allow for tractable policy design. However, first, to complete the formulation of the above problem, we need to characterize the expression for the average AoI, $\Delta^{(N)}_i(\bm{p},\bm{\mu}_1,\bm{\mu}_2)$, which we will derive in the next section. Also, henceforth, we refer to the triple $(p_i,\mu_{1i},\mu_{2i})$ as the policy of device $D_i$.

\section{Age of Information (AoI) Calculations}
In this section, we will now determine the average AoI for device $i$ appearing in Problem \ref{problem:N_user_game}. In this regard, we will use the method of stochastic hybrid systems (SHS) proposed in \cite{hespanha2006modelling} 
first for continuous-state dynamical systems and then in \cite{yates2018age, yates2018status} for discrete-state modeling and AoI computations for different server constellations. For completeness, we first briefly review the main concepts in the next subsection. We refer the interested reader to \cite{yates2018status} for details.
\vspace{-0.5cm}
\subsection{Stochastic Hybrid Systems (SHS)}\label{Subsec:SHS}
\begin{figure}[t]
    \centering
     \includegraphics[width=0.9\columnwidth]{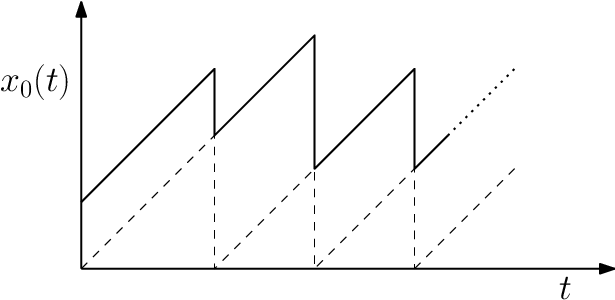}
    \caption{\small{Evolution of AoI at the receiver}}
    \label{fig:Sampe_AoI}
        \vspace{-0.2cm}
\end{figure}

Let us briefly recall how an AoI process as shown in Fig. \ref{fig:Sampe_AoI} can be modeled by a piecewise linear SHS. In \textit{any} SHS, the overall state of the system can be described using a discrete-continuous pair $(s(t),x(t)) \in \operatorname{S} \times \mathbb{R}^{n+1}$ for all time $t \geq 0$, with $\operatorname{S}$ being a finite set. Here, $n+1$ denotes the number of servers involved in the constellation plus the monitor/receiver itself. The continuous state $x(t)$ evolves according to a stochastic differential equation 
\begin{align}
dx(t) = {\tt a}(t,s,x)dt + {\tt b}(t,s,x)dB(t),
\end{align}
where $B(t)$ is a standard Brownian motion, and ${\tt a}$ and ${\tt b}$ are real-valued mappings. Further, the discrete state $s(t)$ evolves from a state $s$ to a state $s'$ with transition intensity ${\tt q}\delta_{s,s'}$ within the set $\operatorname{S}$. The notation $\delta_{s,s'}$ denotes the Kronecker delta function, which equals 1 if and only if $s=s'$, and 0 otherwise. For each such transition, the continuous state also jumps to a new value $x'$ and is defined using the mapping $x' ={\tt h}(t,s,x;s')$. The resulting process $x(t)$ thus has piecewise \textit{continuous} sample paths. 

Let us restrict our attention now to systems in which $s(t)$ evolves as a finite-state Markov chain (FS-MC). Within such a setting, the AoI process can be characterized as a special case of the SHS theory which is a piecewise linear SHS with ${\tt a}(t,s,x) = {\tt u}_s \in \{0,1\}$, $ {\tt b}(t,s,x) = 0 $, and ${\tt h}(t,s,x) = xA_s$, where $A_s \in \{0,1\}^{n+1 \times n+1}$.

Following \cite{yates2018age}, we define $\pi_{s'}(t):= \mathbb{P}(s(t) = s')$ as the discrete-Markov state probability of being at state $s(t) = s'$ and $v_{s'k}(t):= \mathbb{E}[x_k(t) \delta_{s(t) = s'}]$, which measures the correlation between the age process $x_k(t)$ in server $k$ with the discrete state $s(t)$ at timestep $t$. Further, let us denote the set of possible outgoing transitions from a particular state $s$ as $\opL_s:= \{\ell: s_{\ell} = s\}$ and the set of possible incoming transitions to a state $s'$ as $\opL'_{s'}:= \{\ell: s_\ell = s'\}$. Then, under the assumption of ergodicity of the FS-MC, a unique steady state distribution $\Bar{\pi}:= [\bar{\pi}_1, \cdots, \bar{\pi}_m]$ exists \cite{norris1998markov} and satisfies the conservation law,
\begin{subequations}\label{steady_state_prob}
    \begin{align}
    \Bar{\pi}_s \sum_{\ell \in \opL_s}{\tt q}^\ell & = \sum_{\ell' \in \opL'_s} {\tt q}^{\ell'} \Bar{\pi}_{s_{\ell'}}, \quad \forall s \in \opS, \\
    \sum_{s \in \opS} \Bar{\pi}_s &= 1, \label{prob_sum}
\end{align}
\end{subequations}
where $m$ denotes the cardinality of $\operatorname{S}$. Let us denote by $v_s$ the vector $[v_{s0}, v_{s1}, \cdots, v_{sn}]$ of correlations for all servers. We note that here and henceforth, we always take the index of the monitor (which in our case will be the device $D_i$) as 0, which means $x_0$ denotes the age process of the monitor and $v_{s0}$ denotes the correlation function as defined above for the monitor. Consequently, one can obtain the following result.

\begin{theorem}\!\cite[Theorem 4]{yates2018age}\label{Avg_Age_thm}
    Suppose that $\Bar{\pi}$ is the state distribution of the FS-MC and there exists a stationary solution $\Bar{v}\!:=\! [\Bar{v}_1,\! \cdots,\! \Bar{v}_m]$ of the process $v_{\cdot}(t)$ satisfying
    \begin{align}\label{cond_prob_eqn}
        \Bar{v}_s \sum_{\ell \in \opL_s}{\tt q}^\ell = {\tt u}_s \Bar{\pi}_s + \sum_{\ell' \in \opL'_s} {\tt q}^{\ell'} \Bar{v}_{s_{\ell'}}A_{\ell'}.
    \end{align}
    Then, the average AoI is given by $\Delta:= \sum_{s \in \opS}\bar{v}_{s0}$.
\end{theorem}

We will use the above theorem to compute ${\Delta}^{(N)}_i(\bm{p},\bm{\mu_1},\bm{\mu_2})$ as defined in Problem \ref{problem:N_user_game}.

\begin{figure}[t]
    \centering
    \includegraphics[width=0.95\columnwidth]{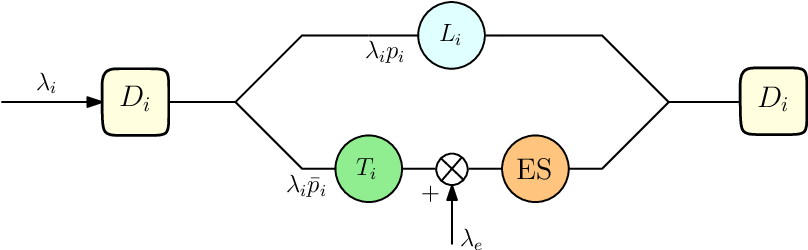}
    \caption{\small{Task flow from the perspective of device $D_i$.}}
    \label{fig:generic_user}
\end{figure}

\subsection{Average AoI Calculation}
In this subsection, we will provide an approximate average AoI expression for device $D_i$ by assuming that the incoming Poisson process has a sufficiently large parameter $\lambda_i, \forall i$. The justification for carrying out an approximate analysis is due to exponential increase in the state space of the FS-MC (defined in the previous subsection), the details of which will be provided later in Remark \ref{rem:Approx_analysis}. Let us begin by concentrating on the incoming task flow from the perspective of device $D_i$ as shown in Fig.~\ref{fig:generic_user}. Under the assumption that the departure process of $T_j$'s for all users closely follows a Poisson process, the exogenous incoming rate (referred to as $\lambda_e$ in Fig. \ref{fig:generic_user}) interfering with the AoI process of $D_i$ approximately follows a $\text{Poi}(\lambda_e)$ distribution \cite{yates2018age}, where $\lambda_e:= \sum_{j=1, j \ne i}^N \frac{\lambda_i\bar{p}_i \mu_{1,i}}{\lambda_i\bar{p}_i + \mu_{1,i}}$. The quantity $\lambda_e$ denotes the cumulative exogeneous rate of users other than the $i$th one.

Next, let us mathematically formulate the AoI process for device $D_i$ using the SHS method as discussed in the previous subsection. To this end, we need to define the state space $\operatorname{S}$ along with its corresponding transition functions to characterize the FS-MC, which is done as follows. Also, henceforth, we will refer to the exogeneous packets as in Fig. \ref{fig:generic_user} as that of class 2.

\textit{1. State Space:} The state space $\opS$ of the FS-MC constitutes 8 states which keep track of the server that is currently servicing the freshest, the second freshest, and the oldest packets of $D_i$ and the packets of class 2. Detailed descriptions are provided in Table \ref{table:states}. For example, state $s_4$ denotes that for device $D_i$, server $T_i$ is idling, and servers $L_i$ and the ES are servicing the freshest and the $2$nd freshest packet of $D_i$, respectively. The choice of the states is made in view of the concept of fake updates \cite{yates2018status}, whereby without any loss of generality, we can assume that all servers \emph{which do not precede a node of packet arrival} ($L_i$ and ES in our case) are busy all the time by running a fake packet in that server. The type and current AoI of the fake packet is the same as the departing packet. Consequently, we have that ${\tt u}_s = [1~1~1~1]$ for $s=s_1,s_2,s_3,s_7,s_8$ and $\opu_s:= \hat {\tt u}_s = [1~0~1~1]$ for $s=s_4,s_5,s_6$ since the transmitter $T_i$ is idling in the latter three states.

\begin{remark}
    It is essential to take care of the emphasized phrase in the later part of the previous paragraph on which server can run a fake packet. This is because the transmitter $T_i$ in our formulation precedes the point of arrival of exogeneous packets (as in Fig. \ref{fig:generic_user}). Thus, the SHS model should take into account whether it is idling or is busy. This prevents us from running a fake update at $T_i$ and we explicitly account for the idle state of $T_i$ by writing ``\textit{no packet}" in Table \ref{table:states}. 
\end{remark}

\begin{table}[t]
    \centering
    \begin{tabular}{ |c|c|c|c| } 
    \hline
    state & server 1 ($T_i$) & server 2 $(L_i)$ & server 3 (ES) \\
    \hline
    $s_1$ & freshest & $2^{nd}$ freshest & oldest \\ \hline
    $s_2$ & freshest & oldest & $2^{nd}$ freshest \\ \hline
    $s_3$ & $2^{nd}$ freshest & freshest & oldest \\ \hline
    $s_4$ & \emph{no packet} & freshest & $2^{nd}$ freshest  \\ \hline
    $s_5$ & \emph{no packet} & $2^{nd}$ freshest & freshest \\ \hline
    $s_6$ & \emph{no packet} & freshest & class 2 \\ \hline
    $s_7$ & freshest & $2^{nd}$ freshest & class 2 \\ \hline
    $s_8$ & $2^{nd}$ freshest & freshest & class 2 \\
    \hline
    \end{tabular}
    \caption{\small{State dictionary for the finite FS-MC}}
    \label{table:states}
\end{table}

\textit{2. Transition Functions:} Now that we have completely described the state state $\opS$, we next list the possible transitions in the FS-MC in Table \ref{table:transitions}. Alongside, we also track the current AoIs of the packets in each server after each transition, which is listed as the AoI vector $x'(t):= [x'_0(t) ~x'_1(t)~x'_2(t)~x'_3(t)]$, where $x'_0(t),$ $x'_1(t),$ $x'_2(t)$, and $x'_3(t)$ denote the AoI at $D_i$, the local processor $L_i$, the transmitter $T_i$ and the ES, respectively. Note that henceforth, we forego the subscript index $i$ for brevity.

\begin{table}[t]
    \centering
    \small
    \begin{tabular}{|c|c|c|c|c|} 
    \hline
    $s$ & ${\tt q}$ & $s'$ & $x' = xA_s$ & $v_sA_s$\\
    \hline
    \multirow{4}{2em}{$s_1$} & $\lambda p$ & $s_3$ & $[x_0 ~x_1 ~0 ~x_3]$ & $[\bar{v}_{10}~\bar{v}_{11}~0~\bar{v}_{13}]$\\
    & $\lambda\Bar{p}$ & $s_1$ & $[x_0 ~0 ~x_2 ~x_3]$ & $[\bar{v}_{10}~0~\bar{v}_{12}~\bar{v}_{13}]$ \\
    & $\lambda_e$ & $s_7$ & $[x_0 ~x_1 ~x_2 ~x_0]$ & $[\bar{v}_{10}~\bar{v}_{11}~\bar{v}_{12}~\bar{v}_{10}]$ \\
    & $\mu_1$ & $s_5$ & $[x_0 ~0 ~x_2 ~x_1]$ & $[\bar{v}_{10}~0~\bar{v}_{12}~\bar{v}_{11}]$ \\
    & $\mu_2$ & $s_1$ & $[x_2 ~x_1 ~x_2 ~x_2]$ & $[\bar{v}_{12}~\bar{v}_{11}~\bar{v}_{12}~\bar{v}_{12}]$ \\
    & $\mu_3$ & $s_1$ & $[x_3 ~x_1 ~x_2 ~x_3]$ & $[\bar{v}_{13}~\bar{v}_{11}~\bar{v}_{12}~\bar{v}_{13}]$ \\ \hline
    \multirow{4}{2em}{$s_2$} & $\lambda p$ & $s_3$ & $[x_0 ~x_1 ~0 ~x_3]$ & $[\bar{v}_{20}~\bar{v}_{21}~0~\bar{v}_{23}]$ \\
    & $\lambda \Bar{p}$ & $s_2$ & $[x_0 ~0 ~x_2 ~x_3]$ & $[\bar{v}_{20}~0~\bar{v}_{22}~\bar{v}_{23}]$ \\
    & $\lambda_e$ & $s_7$ & $[x_0 ~x_1 ~x_2 ~x_0]$ & $[\bar{v}_{20}~\bar{v}_{21}~\bar{v}_{22}~\bar{v}_{20}]$ \\
    & $\mu_1$ & $s_5$ & $[x_0 ~0 ~x_2 ~x_1]$ & $[\bar{v}_{20}~0~\bar{v}_{22}~\bar{v}_{21}]$ \\
    & $\mu_2$ & $s_2$ & $[x_2 ~x_1 ~x_2 ~x_3]$ & $[\bar{v}_{22}~\bar{v}_{21}~\bar{v}_{22}~\bar{v}_{23}]$ \\
    & $\mu_3$ & $s_2$ & $[x_3 ~x_1 ~x_3 ~x_3]$ & $[\bar{v}_{23}~\bar{v}_{21}~\bar{v}_{23}~\bar{v}_{23}]$ \\ \hline
    \multirow{4}{2em}{$s_3$} & $\lambda p$ & $s_3$ & $[x_0 ~x_1 ~0 ~x_3]$ & $[\bar{v}_{30}~\bar{v}_{31}~0~\bar{v}_{33}]$ \\
    & $\lambda \Bar{p}$ & $s_1$ & $[x_0 ~0 ~x_2 ~x_3]$ & $[\bar{v}_{30}~0~\bar{v}_{32}~\bar{v}_{33}]$ \\
    & $\lambda_e$ & $s_8$ & $[x_0 ~x_1 ~x_2 ~x_0]$ & $[\bar{v}_{30}~\bar{v}_{31}~\bar{v}_{32}~\bar{v}_{30}]$ \\
    & $\mu_1$ & $s_4$ & $[x_0 ~0 ~x_2 ~x_1]$ & $[\bar{v}_{30}~0~\bar{v}_{32}~\bar{v}_{31}]$ \\
    & $\mu_2$ & $s_3$ & $[x_2 ~x_2 ~x_2 ~x_2]$ & $[\bar{v}_{32}~\bar{v}_{32}~\bar{v}_{32}~\bar{v}_{32}]$ \\
    & $\mu_3$ & $s_3$ & $[x_3 ~x_1 ~x_2 ~x_3]$ & $[\bar{v}_{33}~\bar{v}_{31}~\bar{v}_{32}~\bar{v}_{33}]$ \\ \hline
    \multirow{4}{2em}{$s_4$} & $\lambda p$ & $s_4$ & $[x_0 ~0 ~0 ~x_3]$ & $[\bar{v}_{40}~0~0~\bar{v}_{43}]$ \\
    & $\lambda\Bar{p}$ & $s_1$ & $[x_0 ~0 ~x_2 ~x_3]$ & $[\bar{v}_{40}~0~\bar{v}_{42}~\bar{v}_{43}]$ \\
    & $\lambda_e$ & $s_6$ & $[x_0 ~0 ~x_2 ~x_0]$ & $[\bar{v}_{40}~0~\bar{v}_{42}~\bar{v}_{40}]$ \\
    & $\mu_2$ & $s_4$ & $[x_2 ~0 ~x_2 ~x_2]$ & $[\bar{v}_{42}~0~\bar{v}_{42}~\bar{v}_{42}]$ \\
    & $\mu_3$ & $s_4$ & $[x_3 ~0 ~x_2 ~x_3]$ & $[\bar{v}_{43}~0~\bar{v}_{42}~\bar{v}_{43}]$ \\ \hline
    \multirow{4}{2em}{$s_5$} & $\lambda p$ & $s_4$ & $[x_0 ~0 ~0 ~x_3]$ & $[\bar{v}_{50}~0~0~\bar{v}_{53}]$ \\
    & $\lambda \Bar{p}$ & $s_2$ & $[x_0 ~0 ~x_2 ~x_3]$ & $[\bar{v}_{50}~0~\bar{v}_{52}~\bar{v}_{53}]$ \\
    & $\lambda_e$ & $s_6$ & $[x_0 ~0 ~x_2 ~x_0]$ & $[\bar{v}_{50}~0~\bar{v}_{52}~\bar{v}_{50}]$ \\
    & $\mu_2$ & $s_5$ & $[x_2 ~0 ~x_2 ~x_3]$ & $[\bar{v}_{52}~0~\bar{v}_{52}~\bar{v}_{53}]$ \\
    & $\mu_3$ & $s_5$ & $[x_3 ~0 ~x_3 ~x_3]$ & $[\bar{v}_{53}~0~\bar{v}_{53}~\bar{v}_{53}]$ \\ \hline
    \multirow{4}{2em}{$s_6$} & $\lambda p$ & $s_6$ & $[x_0 ~0 ~0 ~x_3]$ & $[\bar{v}_{60}~0~0~\bar{v}_{63}]$ \\
    & $\lambda \Bar{p}$ & $s_7$ & $[x_0 ~0 ~x_2 ~x_3]$ & $[\bar{v}_{60}~0~\bar{v}_{62}~\bar{v}_{63}]$ \\
    & $\lambda_e$ & $s_6$ & $[x_0 ~0 ~x_2 ~x_0]$ & $[\bar{v}_{60}~0~\bar{v}_{62}~\bar{v}_{60}]$ \\
    & $\mu_2$ & $s_6$ & $[x_2 ~0 ~x_2 ~x_2]$ & $[\bar{v}_{62}~0~\bar{v}_{62}~\bar{v}_{62}]$ \\
    & $\mu_3$ & $s_6$ & $[x_3 ~0 ~x_2 ~x_3]$ & $[\bar{v}_{63}~0~\bar{v}_{62}~\bar{v}_{63}]$ \\ \hline
    \multirow{4}{2em}{$s_7$} & $\lambda p$ & $s_8$ & $[x_0 ~x_1 ~0 ~x_3]$ & $[\bar{v}_{70}~\bar{v}_{71}~0~\bar{v}_{73}]$ \\
    & $\lambda \Bar{p}$ & $s_7$ & $[x_0 ~0 ~x_2 ~x_3]$ & $[\bar{v}_{70}~0~\bar{v}_{72}~\bar{v}_{73}]$ \\
    & $\lambda_e$ & $s_7$ & $[x_0 ~x_1 ~x_2 ~x_0]$ & $[\bar{v}_{70}~\bar{v}_{71}~\bar{v}_{72}~\bar{v}_{70}]$ \\
    & $\mu_1$ & $s_5$ & $[x_0 ~0 ~x_2 ~x_1]$ & $[\bar{v}_{70}~0~\bar{v}_{72}~\bar{v}_{71}]$ \\
    & $\mu_2$ & $s_7$ & $[x_2 ~x_1 ~x_2 ~x_2]$ & $[\bar{v}_{72}~\bar{v}_{71}~\bar{v}_{72}~\bar{v}_{72}]$ \\
    & $\mu_3$ & $s_7$ & $[x_3 ~x_1 ~x_2 ~x_3]$ & $[\bar{v}_{73}~\bar{v}_{71}~\bar{v}_{72}~\bar{v}_{73}]$ \\ \hline
    \multirow{4}{2em}{$s_8$} & $\lambda p$ & $s_8$ & $[x_0 ~x_1 ~0 ~x_3]$ & $[\bar{v}_{80}~\bar{v}_{81}~0~\bar{v}_{83}]$ \\
    & $\lambda \Bar{p}$ & $s_7$ & $[x_0 ~0 ~x_2 ~x_3]$ & $[\bar{v}_{80}~0~\bar{v}_{82}~\bar{v}_{83}]$ \\
    & $\lambda_e$ & $s_8$ & $[x_0 ~x_1 ~x_2 ~x_0]$ & $[\bar{v}_{80}~\bar{v}_{81}~\bar{v}_{82}~\bar{v}_{80}]$ \\
    & $\mu_1$ & $s_4$ & $[x_0 ~0 ~x_2 ~x_1]$ & $[\bar{v}_{80}~0~\bar{v}_{82}~\bar{v}_{81}]$ \\
    & $\mu_2$ & $s_8$ & $[x_2 ~x_2 ~x_2 ~x_2]$ & $[\bar{v}_{82}~\bar{v}_{82}~\bar{v}_{82}~\bar{v}_{82}]$ \\
    & $\mu_3$ & $s_8$ & $[x_3 ~x_1 ~x_2 ~x_3]$ & $[\bar{v}_{83}~\bar{v}_{81}~\bar{v}_{82}~\bar{v}_{83}]$ \\ \hline
    \end{tabular}
    \caption{\small{State transitions of the FS-MC and associated AoI jumps.}}
    \label{table:transitions}
        \vspace{-0.4cm}
\end{table}

Now that the FS-MC is completely characterized by Tables \ref{table:states} and \ref{table:transitions}, we now proceed towards computing the average AoI for $D_i$. Let us begin by defining the quantities $a:= \lambda + \lambda_e + \mu_1 + \mu_2 + \mu_3$ and $\hat{a} :=a-\mu_1$. Then, using \eqref{steady_state_prob}, the steady state probability vector $\Bar{\pi}$ satisfies \eqref{prob_sum} and the following set of equations in \eqref{eqn_a}-\eqref{eqn_h}:
\begin{subequations}\label{steady_state_prob_this_work}
    \begin{align}
        a \Bar{\pi}_1 & = (\lambda\Bar{p} + \mu_2 + \mu_3)\Bar{\pi}_1 + \lambda\Bar{p} (\Bar{\pi}_3 + \Bar{\pi}_4), \label{eqn_a}\\ 
        a \Bar{\pi}_2 & = (\lambda\Bar{p} + \mu_2 + \mu_3)\Bar{\pi}_2 + \lambda\Bar{p} \Bar{\pi}_5, \\ 
        a \Bar{\pi}_3 & = (\lambda p + \mu_2 + \mu_3)\Bar{\pi}_3 + \lambda p (\Bar{\pi}_1 + \Bar{\pi}_2), \\ 
        \hat{a} \Bar{\pi}_4 & = (\lambda p + \mu_2 + \mu_3)\Bar{\pi}_4 \!+\! \lambda p \Bar{\pi}_5 \!+\! \mu_1 (\Bar{\pi}_3 \!+\! \Bar{\pi}_8), \\ 
        \hat{a} \Bar{\pi}_5 & = (\mu_2 + \mu_3)\Bar{\pi}_5 + \mu_1 (\Bar{\pi}_1 + \Bar{\pi}_2 + \Bar{\pi}_7), \\
        \hat{a} \Bar{\pi}_6 & = (\lambda p + \lambda_e + \mu_2 + \mu_3)\Bar{\pi}_6 + \lambda_e (\Bar{\pi}_4 + \Bar{\pi}_5), \\
        a \Bar{\pi}_7 & = (\lambda\Bar{p} + \lambda_e + \mu_2 + \mu_3)\Bar{\pi}_7 + \lambda_e (\Bar{\pi}_1 + \Bar{\pi}_2)\nonumber\\& ~~~+\! \lambda \Bar{p} (\Bar{\pi}_6 + \Bar{\pi}_8), \\ 
        a \Bar{\pi}_8 & = (\lambda p\! +\! \lambda_e\! \!+ \!\mu_2 \!+\! \mu_3)\Bar{\pi}_7 \!+ \!\lambda_e \Bar{\pi}_3 +\! \lambda p \Bar{\pi}_7. \label{eqn_h} \!\!
    \end{align}
\end{subequations}

\begin{figure*}[h]
\begin{subequations}\label{eqn_v_s}
\begin{small}
    \begin{align}
        a\Bar{v}_1 & \!= \!{\tt u}_s \Bar{\pi}_1 \!+\! \lambda \Bar{p}[\bar{v}_{10}~0~\bar{v}_{12}~\bar{v}_{13}] + \mu_2 [\bar{v}_{12}~\bar{v}_{11}~\bar{v}_{12}~\bar{v}_{12}] + \mu_3 [\bar{v}_{13}~\bar{v}_{11}~\bar{v}_{12}~\bar{v}_{13}] + \lambda \Bar{p} [\bar{v}_{30}~0~\bar{v}_{32}~\bar{v}_{33}]  + \lambda \Bar{p} [\bar{v}_{40}~0~\bar{v}_{42}~\bar{v}_{43}], \label{veqn_a}\\
        a\Bar{v}_2 & = {\tt u}_s \Bar{\pi}_2 + \lambda \Bar{p}[\bar{v}_{20}~0~\bar{v}_{22}~\bar{v}_{23}] + \mu_2 [\bar{v}_{22}~\bar{v}_{21}~\bar{v}_{22}~\bar{v}_{23}] + \mu_3 [\bar{v}_{23}~\bar{v}_{21}~\bar{v}_{23}~\bar{v}_{23}] + \lambda \Bar{p} [\bar{v}_{50}~0~\bar{v}_{52}~\bar{v}_{53}], \\
        a\Bar{v}_3 & \!= \!{\tt u}_s \Bar{\pi}_3 + \lambda p [\bar{v}_{30}~\bar{v}_{31}~0~\bar{v}_{33}] \!+ \!\mu_2 [\bar{v}_{32}~\bar{v}_{32}~\bar{v}_{32}~\bar{v}_{32}] + \mu_3 [\bar{v}_{33}~\bar{v}_{31}~\bar{v}_{32}~\bar{v}_{33}] + \lambda p [\bar{v}_{10}~\bar{v}_{11}~0~\bar{v}_{13}]  + \lambda p [\bar{v}_{20}~\bar{v}_{21}~0~\bar{v}_{23}], \\
        \hat{a}\Bar{v}_4 & = \hat {\tt u}_s \Bar{\pi}_4 + \lambda p ([\bar{v}_{40}~0~0~\bar{v}_{43}] + [\bar{v}_{50}~0~0~\bar{v}_{53}])+ \mu_2 [\bar{v}_{42}~0~\bar{v}_{42}~\bar{v}_{42}] + \mu_3 [\bar{v}_{43}~0~\bar{v}_{42}~\bar{v}_{43}] + \mu_1 ([\bar{v}_{30}~0~\bar{v}_{32}~\bar{v}_{31}] + [\bar{v}_{80}~0~\bar{v}_{82}~\bar{v}_{81}]),\\
        \hat{a}\Bar{v}_5 & = \hat {\tt u}_s \Bar{\pi}_5 + \mu_1 [\bar{v}_{70}~0~\bar{v}_{72}~\bar{v}_{71}] + \mu_2 [\bar{v}_{52}~0~\bar{v}_{52}~\bar{v}_{53}] + \mu_3 [\bar{v}_{53}~0~\bar{v}_{53}~\bar{v}_{53}] + \mu_1 [\bar{v}_{10}~0~\bar{v}_{12}~\bar{v}_{11}] + \mu_1 [\bar{v}_{20}~0~\bar{v}_{22}~\bar{v}_{21}], \\
        \hat{a}\Bar{v}_6 & = \hat {\tt u}_s \Bar{\pi}_6 + \lambda p [\bar{v}_{60}~0~0~\bar{v}_{63}] + \mu_2 [\bar{v}_{62}~0~\bar{v}_{62}~\bar{v}_{63}] + \mu_3 [\bar{v}_{63}~0~\bar{v}_{62}~\bar{v}_{63}] + \lambda_e ([\bar{v}_{40}~0~\bar{v}_{42}~\bar{v}_{40}] +[\bar{v}_{50}~0~\bar{v}_{52}~\bar{v}_{50}] +[\bar{v}_{60}~0~\bar{v}_{62}~\bar{v}_{60}]), \\
        a\Bar{v}_7 & = {\tt u}_s \Bar{\pi}_7 + \lambda \Bar{p}[\bar{v}_{60}~0~\bar{v}_{62}~\bar{v}_{63}] + \lambda \Bar{p}[\bar{v}_{70}~0~\bar{v}_{72}~\bar{v}_{73}] + \lambda \Bar{p}[\bar{v}_{80}~0~\bar{v}_{82}~\bar{v}_{83}] + \lambda_e [\bar{v}_{10}~\bar{v}_{11}~\bar{v}_{12}~\bar{v}_{10}] + \lambda_e [\bar{v}_{20}~\bar{v}_{21}~\bar{v}_{22}~\bar{v}_{20}] \nonumber \\
        & ~~ + \lambda_e [\bar{v}_{70}~\bar{v}_{71}~\bar{v}_{72}~\bar{v}_{70}] + \mu_3 [\bar{v}_{73}~\bar{v}_{71}~\bar{v}_{72}~\bar{v}_{73}] + \mu_2 [\bar{v}_{72}~\bar{v}_{71}~\bar{v}_{72}~\bar{v}_{72}], \\
        a\Bar{v}_8 & = {\tt u}_s \Bar{\pi}_8 \!+\! \lambda p ([\bar{v}_{70}~\bar{v}_{71}~0~\bar{v}_{73}] \!+\! [\bar{v}_{80}~\bar{v}_{81}0\bar{v}_{83}]) \!+\! \lambda_e ([\bar{v}_{30}~\bar{v}_{31}~\bar{v}_{32}~\bar{v}_{30}] \!+\! [\bar{v}_{80}~\bar{v}_{81}~\bar{v}_{82}~\bar{v}_{80}])\!+\! \mu_2 [\bar{v}_{82}~\bar{v}_{82}~\bar{v}_{82}~\bar{v}_{82}] \!+\! \mu_3 [\bar{v}_{83}~\bar{v}_{81}~\bar{v}_{82}~\bar{v}_{73}].\label{veqn_h}
    \end{align}
    \vspace{-2mm}
    \end{small}
    \hrule
    \vspace{-5mm}
\end{subequations}
\end{figure*}

The above set of linear equations \eqref{steady_state_prob_this_work} allows us to compute the distribution $\bar \pi$ by combining with \eqref{prob_sum}. Then, to compute the average AoI $\Delta^{(N)}_i(\bm{p},\bm{\mu}_1,\bm{\mu}_2)$, it remains to compute the steady state vector $\bar v$ in Theorem \ref{Avg_Age_thm} and then apply the formula for $\Delta$ in the theorem. To compute $\bar v$, we use \eqref{cond_prob_eqn} in Theorem \ref{Avg_Age_thm} to write down the set of linear equations satisfied by its components in \eqref{eqn_v_s}. Consequently, the average AoI, $\Delta^{(N)}_i(\bm{p},\bm{\mu}_1,\bm{\mu}_2)$, can be computed by first computing $\bar \pi$ using \eqref{eqn_a}-\eqref{eqn_h}, substituting the same in \eqref{veqn_a}-\eqref{veqn_h} and solving the latter set of equations. We summarize the result in the following theorem by resuming the use of subscript $i$ corresponding to device $D_i$.

\begin{theorem}\label{thm:AoI_MEC_equitable}
    Suppose that the arrival rate at device $D_i$ is distributed as $\text{Poi}(\lambda_i)$ and its service rates as $\text{exp}(\mu_{1i})$ and $\text{exp}(\mu_{2i})$. Let the service rate of the ES be distributed as $\text{exp}(\mu_3)$. Then, the average AoI ${\Delta}^{(N)}_i(\bm{p},\bm{\mu_1},\bm{\mu_2})$ exists and is given by solving \eqref{steady_state_prob_this_work} and \eqref{eqn_v_s}.
\end{theorem}

We next note here that even though one can obtain the closed-form expressions for the average AoI using Theorem \ref{thm:AoI_MEC_equitable}, long expressions preclude us from writing them in the paper. Furthermore, later, we will provide an algorithm to compute the equilibrium policies for all the devices, where we would not require symbolic expressions of the AoI but only a ``function call'' to solve a linear program, which can then be conveniently solved using any linear program solver.

We have thus completed the average AoI calculations, and now state the following important remark on how the approximate calculations performed in this subsection lead to tractable computations of the average AoI in the presence of a large number of users.

\begin{figure}[t]
    \centering
    \includegraphics[width=0.90\columnwidth]{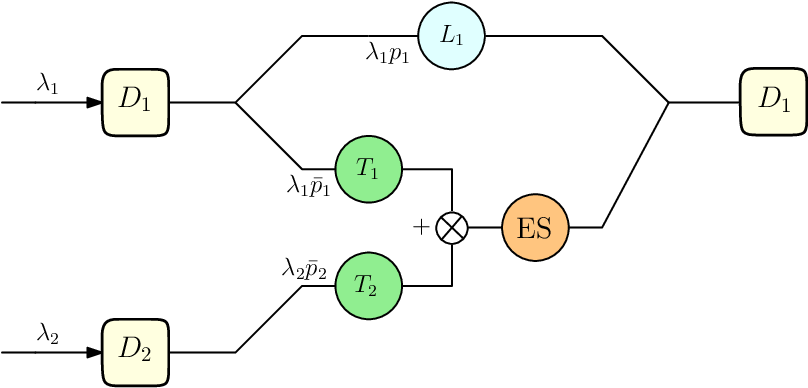}
    \caption{\small{Task flow for a two-user MEC system with one ES from the perspective of device $D_1$.}}
    \label{fig:2_user}
        \vspace{-0.2cm}
\end{figure}

\begin{remark}\label{rem:Approx_analysis}
To understand the importance of the large $\lambda$ assumption, let us focus, for simplicity, on a two-user case in Fig. \ref{fig:N_user} and the perspective of device $D_1$ as shown in Fig. \ref{fig:2_user}. Using this, one can perform the exact AoI computations since the incoming arrivals of both packets at devices $D_1$ and $D_2$ are Poisson, and hence, we can apply the SHS theory without any assumption on $\lambda_i$. In Fig. \ref{fig:generic_user}, however, this is not the case since the exogenous process preceding the ES is not Poisson, which makes the underlying finite state evolution of $s(t)$ non-Markovian and hence, precludes the application of the SHS theory. However, under the large $\lambda$ approximation, the exogenous process closely follows a Poisson process, in which case, we can readily apply the SHS theory. Now, one could ask the question as to why not proceed with the set-up of Fig. \ref{fig:2_user}. The reason for that is the exponential explosion of the state space of the underlying FS-MC with the number of users $N$. It is easy to see that one requires $2^{N-1}$ times the number of states currently required (which is 8) to characterize the FS-MC completely because one would need to track whether the transmitter $T_j$ of the other devices is busy or idle. Thus, approximate analysis (as we carried out with large $\lambda_i$'s and summarized in Theorem \ref{thm:AoI_MEC_equitable}) allows us to compute tractable AoI expressions with the state space being independent of the number of users.
\end{remark}

With the above computations, we have now completely characterized Problem \ref{problem:N_user_game}. In the next section, we proceed toward computing completely distributed Nash equilibrium (NE) solutions to Problem \ref{problem:N_user_game} using the MFG approach.

\section{Mean-Field Game (MFG)}
Now that we have completely characterized the non-cooperative game problem defined in Problem \ref{problem:N_user_game}, our aim is to compute NE policies $(p_i^*,\mu_{1i}^*, \mu_{2i}^*), \forall i$, which satisfy the following set of inequalities \cite{bacsar1998dynamic}, 
\begin{align}
    & J^{N,i}(p_i^*,\mu_{1i}^*, \mu_{2i}^*,p_{-i}^*,\mu_{1,-i}^*, \mu_{2,-i}^*) \nonumber\\
    & \qquad \leq J^{N,i}(p_i,\mu_{1i},\mu_{2i},p_{-i}^*,\mu_{1,-i}^*, \mu_{2,-i}^*),~\forall i \in [N],
\end{align}
where the notation $x_{-i}$ stands for the vector of variables $x_j$ for all users $j$ excluding user $i$. Briefly, the above set of inequalities state that any \textit{rational} user who tries to deviate from the NE policy $(\bm{p}^*,\bm{\mu_1}^*,\bm{\mu_2}^*)$ incurs a higher cost, and thus it is in the best interest of each user to follow the NE policy. However, since the cost function of each user depends on the policies of the other users, no user can independently optimize to compute its own policy. Hence, each user requires the knowledge of the policies of the others in the population, which (a) can be difficult to acquire particularly within a large user setting, and (b) can incur significant communication overhead. Thus, to alleviate this issue, we would like to design completely distributed NE policies for the users where \textit{each of them utilizes only their own local information} and the statistical information of the system (such as the limiting distribution $\mathbb P(\phi)_{\forall \phi \in \Phi}$). In this regard, we leverage the elegant framework of MFGs \cite{al2015joint, huang2007large, lasry2007mean}. 

Motivated by statistical physics, the theory of MFGs aims to approximate the complex interactions within an interacting particle system with an \textit{average effective field} (also referred to as the mean-field) generated by the corresponding infinite particle system. under suitable assumptions of ``symmetry'', "indistinguishability",  and ``anonymity''. This creates a \textit{decoupling} effect and each particle now best reacts to the \textit{mean-field} generated by the entire population in a manner such that its own behavior is consistent with that of the other particles without worrying about the policies of other particles within the population, where ``consistency" is ensured through the solution of a fixed-point equation. Consequently, to derive equilibrium solutions, it suffices to consider the perspective of one generic particle, which represents the entire population. With the above prelude on MFGs, let us now set up the generic device's optimization problem in the next subsection.

\subsection{Generic Device Optimization Problem}
Let us begin by focusing on a generic device of type $\phi \in \Phi$. The tasks arriving at the generic device $D_\phi$ are distributed as a $\text{Poi}(\lambda_\phi)$ r.v. which are sent to either its local processor or to the transmitter for offloading by employing a mean $p_\phi$ i.i.d. Bernoulli distributed r.v. This divides the incoming arrival process into two independent Poisson processes with respective means $\lambda_\phi p_\phi$ and $\lambda_\phi \Bar{p}_\phi$. In addition, the service times of the generic transmitter and the generic local processor are distributed as exponential r.v.s with parameters $\mu_{1,\phi}$ and $\mu_{2,\phi}$, respectively. The corresponding upper bounds on the service rates are denoted as $P_{\phi,max}$ and $f_{\phi,max}$, respectively.

Next, let us introduce the following quantities $\rho^{(N)}  := \frac{\lambda_e}{N \mu_3}$ and $\rho := \lim_{N \rightarrow \infty}\rho^{(N)}$, where $\rho^{(N)}$ denotes the mean load on the ES and $\rho$ denotes the infinite user (or the MF) approximation as discussed earlier. Then, we have that for a large user MEC system, the exogenous arrival rate $\lambda_e$ can be approximated as
\begin{align}
    \lambda_e = (N-1) \mu_3 \times \frac{\lambda_e}{ (N-1) \mu_3}\approx (N-1) \mu_3 \rho. 
\end{align}
Consequently, the average AoI of the generic device becomes
\begin{align}\label{MF_AoI}
    \!\!\!{\Delta}_\phi(p_\phi,\mu_{1,\phi},\mu_{2,\phi},\rho)\!:=\! \Delta^{(N)}_{\phi_i}(\bm{p},\bm{\mu_1},\bm{\mu_2},\lambda_e)\!\mid_{\lambda_e = (N\!-\!1) \rho \mu_3},\!\!
\end{align}
where the notation $x(z)\mid_{z = a}$ denotes the value of $x$ when $a$ is substituted for the argument $z$. Then, we can formally state the generic device's optimization problem as follows.

\begin{problem}[Generic device optimization problem]\label{problem:generic_user}
    \begin{align*}
        \min_{(p_\phi,\mu_{1,\phi},\mu_{2,\phi}) \in [0,1] \times \mathbb{R}^2} &  J_{\rho}(p_\phi,\mu_{1,\phi},\mu_{2,\phi}) \\
        &  \hspace{-16mm}\mbox{s.t.}~~ ~ 0\leq \mu_{1,\phi} \leq P_{\phi,max} \\
        & \hspace{-9mm}0\leq \mu_{2,\phi} \leq f_{\phi,max},
        \end{align*} 
        where $J_{\rho}(p_\phi,\mu_{1,\phi},\mu_{2,\phi})  := t_{T_\phi}\mu_{1,\phi} + t_{L_\phi}\eta_\phi \mu_{2,\phi}^3 + V_\phi {\Delta}_\phi(p_\phi,\mu_{1,\phi},\mu_{2,\phi},\rho),$ with 
$t_{L_\phi} = \nicefrac{\lambda_\phi p_\phi}{(\lambda_\phi p_\phi + \mu_{2,\phi})}$, $t_{T_\phi} = \nicefrac{\lambda_\phi \bar p_\phi}{(\lambda_\phi \bar p_\phi + \mu_{1,\phi})}$ and ${\Delta}_\phi(p_\phi,\mu_{1,\phi},\mu_{2,\phi},\rho)$ is defined in \eqref{MF_AoI}.
\end{problem}

Consequently, the MFG is defined using two operators, namely the optimality and the consistency operators as:
\begin{enumerate}
    \item Optimality for generic user of type $\phi, \forall \phi \in \Phi$: 
    \begin{align*}
        (\hat{p}_\phi,\hat{\mu}_{1,\phi},\hat{\mu}_{2,\phi}) & = \Psi_1(\rho) := \text{argmin} ~J_\rho(p_\phi,\mu_{1,\phi},\mu_{2,\phi})
    \end{align*}
    subject to the constraints in Problem \ref{problem:generic_user}.
    \item Consistency of the mean-field: 
    
    $\!\!\hat \rho  = \Psi_2(\hat{p}_\phi,\hat{\mu}_{1,\phi},\hat{\mu}_{2,\phi}) := \frac{1}{\mu_3}\mathbb{E}_{\mathbb P(\phi)}\Big[\frac{\lambda_\phi \hat{\Bar{p}}_\phi \hat{\mu}_{1,\phi}}{\lambda_\phi \hat{\Bar{p}}_\phi + \hat{\mu}_{1,\phi}}\Big].$
\end{enumerate}

Briefly, the optimality operator $\Psi_1(\cdot)$ outputs an optimal policy ($p_\phi,\mu_{1,\phi},\mu_{2,\phi}$) for a user of type $\phi$ for a \textit{given} value of $\rho$. Furthermore, the consistency operator $\Psi_2(\cdot)$ generates a new $\rho$ by using the above optimal policy ($p_\phi,\mu_{1,\phi},\mu_{2,\phi}$) and captures the fact that the generic user should act in a way such that its behavior is \textit{consistent} with the load generated at the ES. The mean-field equilibrium (MFE) which constitutes the tuple of equilibrium policies for all types $(p_{\phi,\text{MFE}},\mu_{1,\phi,\text{MFE}},\mu_{2,\phi,\text{MFE}})_{\forall \phi}$ and the equilibrium mean-field ($\rho_{\text{MFE}}$), is given by the fixed point of the composite map of $\Psi_1$ and $\Psi_2$. To compute the MFE of the MEC game, we next provide a low complexity Algorithm \ref{alg:MFE}, which is based on the technique of projected block-coordinate gradient-descent. 

\begin{algorithm}[t]
	\caption{Fixed-point iteration for computing MFE policy of a generic device}
 \label{alg:MFE}
	\begin{algorithmic}[1]
        \STATE {\textbf{Input:} $V_\phi,\eta_\phi,\lambda_\phi,\mu_3, ~\forall \phi$ \hfill\# System parameters}
        \STATE {\textbf{Input:} $\varepsilon_1, \varepsilon_2$ \hfill \# tolerance parameters}
        \STATE {\textbf{Input:} $\gamma_1,\gamma_2$ \hfill \# Iteration step sizes}
		\STATE {Initialize: $\hat{\rho}^{(0)}$, $\sigma_\phi^{(0)}:= (p_\phi^{(0)},\mu_{1,\phi}^{(0)},\mu_{2,\phi}^{(0)}), \forall \phi$}
        \STATE {$k \gets 1$}
        \WHILE{$|\hat{\rho}^{(m)} - \hat{\rho}^{(m-1)}| > \varepsilon_1$}
        \FOR{$\phi \in \Phi$}
        \STATE {$\ell \gets 1$}
        \WHILE{$|\hat{\sigma}_\phi^{(m')}(1) - \hat{\sigma}_\phi^{(m'-1)}(1)| > \varepsilon_2$}
        \STATE $\hat{\sigma}_\phi^{(\ell)}\!(1)\!\leftarrow \!\hat{\sigma}_\phi^{(\ell-1)}\!(1) \!-\! \gamma_2\nabla \! J_{\hat{\rho}^{(k\!-\!1)}}(p_\phi^{(\ell\!-\!1)}\!\!,\mu_{1,\phi}^{(k\!-\!1)}\!\!,\mu_{2,\phi}^{(k\!-\!1)})$
        \STATE{$\ell \gets \ell + 1$}
        \ENDWHILE
        \STATE $p_\phi^{(k)} = \hat \sigma_\phi^{(m')}(1)$
        \STATE {$\ell \gets 1$}
        \WHILE{$|\hat{\sigma}_\phi^{(m')}(2) - \hat{\sigma}_\phi^{(m'-1)}(2)| > \varepsilon_2$}
        \STATE $\hat{\sigma}_\phi^{(\ell)}(2)\!\leftarrow \! \hat{\sigma}_\phi^{(\ell-1)}(2) \!- \!\gamma_2\nabla \!J_{\hat{\rho}^{(k\!-\!1)}}(p_\phi^{(k)}\!,\mu_{1,\phi}^{(\ell\!-\!1)}\!,\mu_{2,\phi}^{(k\!-\!1)})$
        \STATE{$\ell \gets \ell + 1$}
        \ENDWHILE
        \STATE $\mu_{1,\phi}^{(k)} = \hat \sigma_\phi^{(m')}(2)$
        \STATE {$\ell \gets 1$}
        \WHILE{$|\hat{\sigma}_\phi^{(m')}(3) - \hat{\sigma}_\phi^{(m'-1)}(3)| > \varepsilon_2$}
        \STATE $\hat{\sigma}_\phi^{(\ell)}(3)\leftarrow \hat{\sigma}_\phi^{(\ell-1)}(3) - \gamma_2\nabla J_{\hat{\rho}^{(k\!-\!1)}}(p_\phi^{(k)},\mu_{1,\phi}^{(k)},\mu_{2,\phi}^{(\ell\!-\!1)})$
        \STATE{$\ell \gets \ell + 1$}
        \ENDWHILE
        \STATE $\mu_{2,\phi}^{(k)} = \hat \sigma_\phi^{(m')}(3)$
        \ENDFOR
        \STATE $\hat{\rho}^{(k)} \leftarrow (1-\gamma_1)\hat{\rho}^{(k-1)} + \gamma_1 \mathbb{E}_{\mathbb P(\phi)} \Big[\frac{\lambda_\phi(1-\hat{p}^{(k)}_\phi)\hat{\mu}_{1,\phi}^{(k)}}{\mu_3(\lambda_\phi(1-\hat{p}^{(k)}_\phi)+\hat{\mu}_{1,\phi}^{(k)})}\Big]$
        \STATE {$k \gets k+1$}
        \ENDWHILE
        \STATE \textbf{Output: } $\hat{\rho}^{(m)}, \sigma_\phi^{(m)}, ~\forall \phi.$
	\end{algorithmic}
\end{algorithm}

In summary, Algorithm \ref{alg:MFE} computes a fixed point of the composite operator $\Psi_2 \circ \Psi_1$. Thus, we start by randomly initialization the policy and the mean-field (as in line 4 of the algorithm). Then, given a value of $\rho$, we solve all generic users' optimization problems defined in Problem \ref{problem:generic_user} using block coordinate gradient descent method \cite{wright2015coordinate} (lines 8-15). Subsequently, we update the mean-field via Krasnoselskij's iteration using the optimal policy obtained (line 27). The MFE is then given by the final iterate (line 30). We also note here that the main advantage of the above proposed MFG approach is that once we compute the MF $\rho$ offline, when the users operate in real-time, they would be able to make decisions based on only local information and the pre-computed MF. This shows the fully distributed nature of the MF approach as opposed to the NE based approaches in the literature to design offloading policies.

We also note that a simple computation of the Hessian of $J_\rho$ reveals its highly non-convex nature in the policy of the device of type $\phi$, which makes the resulting MFG, a non-convex game. This means that there could possibly exist multiple MFGs for the aforementioned game. Thus, in search of the best one, we run Algorithm \ref{alg:MFE} for multiple random initializations, and subsequently pick the best (in the case where the resultant MFEs are comparable). Finally, to see how the MFG approach which is based on the infinite-user approximation of the finite-user system performs, we also provide Algorithm \ref{alg:Nash_game} to compute an NE for the MEC game. We will use this later to carry out extensive performance evaluation of the proposed MFG approach to demonstrate that the MFG approximates the $N$-user Nash game reasonably well.

This concludes the analysis of the offloading policy design for the MEC game with equitable access to all the users. In the next section, we will formulate the MEC system consisting of primary and secondary users and will compute NE policies.

\begin{algorithm}[t]
    \caption{Best response Dynamics for computing a Nash equilibrium policy}
    \label{alg:Nash_game}
    
    \begin{algorithmic}[1]
    \STATE {\textbf{Input:} $V_j,\eta,\lambda_j,\mu_3, ~\forall j \in [N]$ \hfill\# System parameters}
    \STATE {\textbf{Input:} $\varepsilon_1,\varepsilon_2$ \hfill \# tolerance parameter}
    \STATE {\textbf{Input:} $\gamma$ \hfill \# Step size of gradient descent}
    \STATE {Initialize:  $\sigma_\phi^{(0)}:= (p_j^{(0)},\mu_{1j}^{(0)},\mu_{2j}^{(0)}), \forall j \in [N]$}
    \STATE {$k \gets 1$}
    \WHILE{$|\hat{\sigma}^{(m)} - \hat{\sigma}^{(m-1)}| > \varepsilon_1$}
    \FOR{$j \in [N]$}
    \STATE {$\ell \gets 1$}
    \WHILE{$|\hat{\sigma}_j^{(m')}(1) - \hat{\sigma}_j^{(m'-1)}(1)| > \varepsilon_2$}
    \STATE $\hat{\sigma}_j^{(\ell)}\!(1)\!\leftarrow \!\hat{\sigma}_j^{(\ell-1)}\!(1) \!-\! \gamma\nabla \! J_{\sigma_{-j}^{(k\!-\!1)}}(p_j^{(\ell\!-\!1)}\!\!,\mu_{1j}^{(k\!-\!1)}\!\!,\mu_{2j}^{(k\!-\!1)})$
    \STATE {$\ell \gets \ell+1$}
    \ENDWHILE
    \STATE $p_j^{(k)} = \hat \sigma_j^{(m')}(1)$
    \STATE {$\ell \gets 1$}
    \WHILE{$|\hat{\sigma}_j^{(m')}(2) - \hat{\sigma}_j^{(m'-1)}(2)| > \varepsilon_2$}
    \STATE $\hat{\sigma}_j^{(\ell)}(2)\!\leftarrow \! \hat{\sigma}_j^{(\ell-1)}(2) \!- \!\gamma\nabla \!J_{\sigma_{-j}^{(k\!-\!1)}}(p_j^{(k)}\!,\mu_{1j}^{(\ell\!-\!1)}\!,\mu_{2j}^{(k\!-\!1)})$
    \STATE {$\ell \gets \ell+1$}
    \ENDWHILE
    \STATE $\mu_{1j}^{(k)} = \hat \sigma_j^{(m')}(2)$
    \STATE {$\ell \gets 1$}
    \WHILE{$|\hat{\sigma}_j^{(m')}(3) - \hat{\sigma}_j^{(m'-1)}(3)| > \varepsilon_2$}
    \STATE $\hat{\sigma}_j^{(\ell)}(3)\leftarrow \hat{\sigma}_j^{(\ell-1)}(3) - \gamma\nabla J_{\sigma_{-j}^{(k\!-\!1)}}(p_j^{(k)},\mu_{1j}^{(k)},\mu_{2j}^{(\ell\!-\!1)})$
    \STATE {$\ell \gets \ell+1$}
    \ENDWHILE
    \STATE $\mu_{2j}^{(k)} = \hat \sigma_j^{(m')}(3)$
    \ENDFOR
    \STATE {$k \gets k+1$}
    \ENDWHILE
    \STATE \textbf{Output: } $(\hat{p}^{(m)}_\phi, \hat{\mu}_{1j}^{(m)},\hat{\mu}_{2j}^{(m)}), ~\forall j \in [N].$
    \end{algorithmic}
\end{algorithm}

\section{MEC With Priority-Based Access}
\begin{figure}[t]    \centerline{\includegraphics[width=0.9\columnwidth]{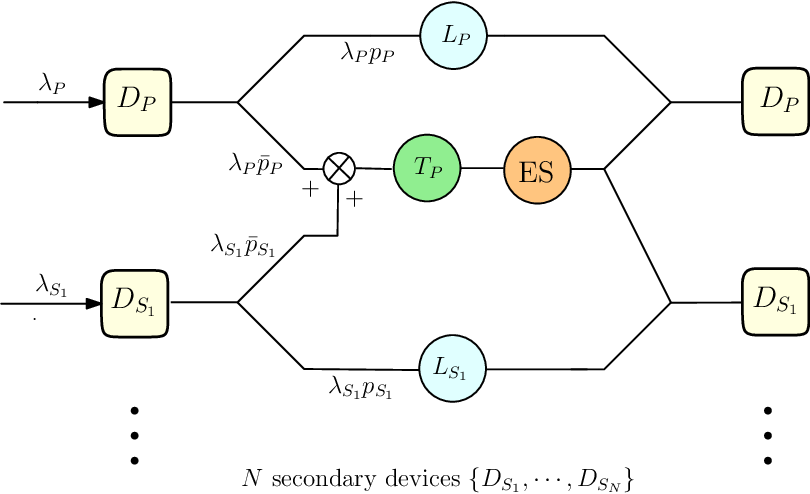}}
	\caption{\small{Task flow for the primary device $D_P$ and the $ith$ secondary device $D_{S_i}$.}}\label{fig:N_user_major_minor}
\end{figure}

Beginning with this section, we extend our earlier MEC framework (with equitable access) to an advanced setting comprised of one primary user and $N$ secondary users as shown in Fig.~\ref{fig:N_user_major_minor}. The primary user has priority access to the ES and the secondary users can offload computations to the ES by utilizing the transmitter of the primary user. The setting is deeply motivated by the \textit{cognitive radio network} technology which aims to promote better effective utilization of the spectrum---a licensed resource. Within the same framework, the secondary users have limited access to the spectrum (which is originally reserved for the primary user) and are allowed to use it only when the primary user is inactive/idling. In this work, we propose a similar model for the MEC system\footnote{Although, we consider only one ES here, the techniques employed can be generalized to more complex settings with multiple ESs which are connected to the secondary users on a graph network. This setting lies within the realm of device-to-device/peer-to-peer communication in computer networks and oligopoly markets in economics and constitutes a promising future research direction.} wherein the primary user has priority access to a computing facility (which we also call the ES to maintain consistency with earlier sections) and secondary users can access  the ES, but with a lower priority. We will formalize the priority discipline in a later section. 

Our objective in this setup is again to solve for the NE policies for both the primary and the secondary users. However, the presence of the primary user leads to an interesting observation here compared to the setup of the previous sections. In the equitable access-MEC setup, all the users have a \emph{vanishing} effect on the mean-field, i.e., the mean-field remains unaffected if finitely many users leave the system or deviate by behaving \textit{irrationally}. In the current setup, however, the primary user (as we will also see later) has a non-vanishing effect. To handle this, we will leverage the paradigm of major-minor mean-field games (MM-MFGs) from control theory literature \cite{nourian2013,firoozi2017execution}. The idea is derived from banking systems where a \textit{finite} number of ``major'' banks (or major players) can significantly affect the operations of ``minor'' banks (or minor players); however, the ``minor'' ones can only affect the operations of the ``major'' ones aggregatively (i.e., via their mean-field). With the above prelude to MM-MFGs, let us proceed with the formulation of the problem.

Consider a primary device $D_P$ which can process the incoming tasks on either its local processor $L_P$ or offload it to the ES using its transmitter $T_P$. Incoming tasks are distributed as a $\text{Poi}(\lambda_P)$ r.v. and are split for local processing or offloading using a Bernoulli r.v. with parameter $p_P$ as shown in Fig.~\ref{fig:N_user_major_minor}. The service time of the local processor and the transmitter of $D_P$ are distributed as $\text{exp}(\mu_{2P})$ and $\text{exp}(P\mu_{1P})$ r.v.s., respectively, with the frequency parameter upper bounded as $\mu_{2P} \leq f_{P,max}$ and the power consumption parameter upper bounded as $\mu_{1P} \leq P_{max}$. The associated power consumed at the local processor is $\eta_P \mu_{2P}^3$ with $\eta_P > 0$. 

Next, the parameters and service notions of each secondary user $D_{S_i}$ and the ES are defined in the same way as for the MEC system of Section~\ref{sec:MEC_formulation}, and hence, are not repeated here. We follow the \textit{last-come-first-serve with priority preemption} (LCFS-PP) discipline, wherein if a primary device's packet is being served at any server, any incoming secondary device's packet is dropped and not allowed to be served. Only another packet of the primary user can preempt it. Secondary users however, get equitable access to both $T_P$ and the ES and any secondary user can preempt the packet of any other secondary user. Both the primary and secondary users aim to minimize their power consumptions and average AoI of their offloaded packets. Additionally, in lieu of allowing a secondary user to access its transmitter, the primary user charges the secondary users an amount proportional to the loading created at $T_P$, which we will formalize next. 

In contrast to the case of the MEC system with equitable access where there was no primary user, in this case, the transmitter of $T_P$ is utilized by the secondary devices as well. Thus, the power expended for offloading is due to both its own packets and those of the secondary users whenever allowed. Thus, the fraction of time that the $T_P$ is busy serving \textit{its own packets} can be computed as 
\begin{align}
    t_{T_P,1} = \frac{\lambda_P  \bar p_P}{(\lambda_P  \bar p_P + \mu_{1P})}.
\end{align}
Further, the fraction of time that the transmitter is busy serving packets of the secondary users can be computed as 
\begin{align}
    t_{T_P,2} & =   \frac{\sum_{j=1}^N\lambda_j \bar p_j}{\sum_{j=1}^N \lambda_j \bar p_j + (\mu_{1P} + \lambda_P  \bar p_P)} \frac{\mu_{1P}}{\mu_{1P} + \lambda_P  \bar p_P},
    \end{align}
where the first multiplying fraction in the above expression denotes the average busy period of $T_P$ serving the exogeneous secondary input of $\lambda_s := \sum_{j=1}^N\lambda_j \bar p_j$ and the second one denotes the fraction of the time that the $T_P$ is not serving the packets of the primary user. Subsequently, the mean service time spent by $T_P$ in task offloading is obtained as $ t_{T_P} = t_{T_P,1} + t_{T_P,2}.$ In addition, the fraction of time that the primary device's local processor is busy can be computed as 
\begin{align}
t_{L_P} = \frac{\lambda_P  p_P}{(\lambda_P  p_P + \mu_{2P})}.
\end{align}

Thus, we can formally state the primary and the secondary devices' optimization problems as below.

\begin{problem}[Primary device's optimization problem]\label{problem:primary_user}
    \begin{align*}
        \min_{(p_P,\mu_{1P},\mu_{2P}) \in [0,1] \times \mathbb{R}^2} &  J_P^{(N)}(p_P,\mu_{1P},\mu_{2P},\lambda_s) \\
        &  \hspace{-16mm}\mbox{s.t.}~~ ~ 0\leq \mu_{1P} \leq P_{max} \\
        & \hspace{-9mm} 0\leq \mu_{2P} \leq f_{P,max},
    \end{align*} 
    where $J_P^{(N)}(p_P,\mu_{1P},\mu_{2P},\lambda_s)  := t_{T_P}\mu_{1P} + t_{L_P}\eta_P \mu_{2P}^3 + V_P {\Delta}_P(p_P,\mu_{1P},\mu_{2P}) - \alpha t_{T_P,2}$. Here, ${\Delta}_P(p_P,\mu_{1P},\mu_{2P})$ denotes the average AoI incurred by the primary user and $\alpha \geq 0$ denotes the fixed price charged by the primary user for using its transmitter. Following this, the overall reward received $\alpha t_{T_P,2}$ from serving secondary users is proportional to the fraction of time that it serves secondary users.
\end{problem}

\begin{figure}[t]
    \centerline{\includegraphics[width=0.9\columnwidth]{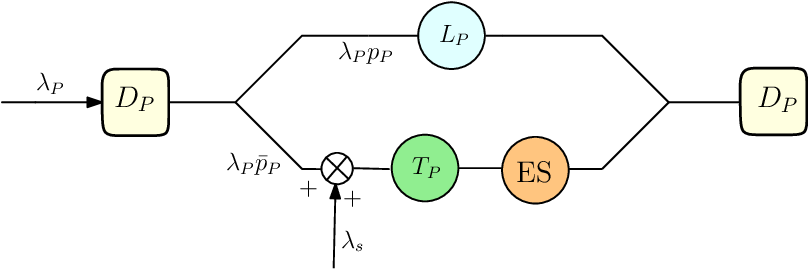}}
	\caption{\small{Task flow for primary device with exogeneous incoming $\lambda_s$.}}
	\label{Fig:N_user_major_minor_single_major}
\end{figure}

Next, let us define the joint decision variable vectors as $\boldsymbol{p_{ps}}:= [p_1, \cdots, p_N, p_P]$ and $\boldsymbol{\mu_{2,ps}}:= [\mu_{21}, \cdots, \mu_{2N}, \mu_{2P}]$, and the average AoI incurred by a secondary user as $\Delta^{(N)}_i(\bm{p_{ps}}, \bm{\mu_{2,ps}},\bm{\mu_{1P}})$. Then, the secondary device's optimization problem is stated as follows.

\begin{problem}[Secondary device's optimization problem]\label{problem:sec_user}
    \begin{align*}
        \min_{(p_i,\mu_{2i}) \in [0,1] \times \mathbb{R}} &  J_{S_i}^{(N)}(\bm{p_{ps}},\bm{\mu_{2,ps}}, \bm{\mu_{1P}}) \\
        & \hspace{-9mm}\mbox{s.t.}~~ ~ 0\leq \mu_{2i} \leq f_{i,max},
    \end{align*} 
    where $\!J_{S_i}^{(N)}(\bm{p_{ps}},\bm{\mu_{2,ps}}, \bm{\mu_{1P}}) \!\! :=\! \!t_{L_i}\eta_i \mu_{2i}^3 + V_i \Delta^{(N)}_i(\bm{p_{ps}}$ $,\bm{\mu_{2,ps}}, \bm{\mu_{1P}})+ \alpha t_{T_P,2} \lambda_i \bar p_i$, where we recall that $t_{L_i} = \nicefrac{\lambda_i p_i}{(\lambda_i p_i + \mu_{2i})}$ denotes the fraction of time that the local processor $L_{S_i}$ is busy. Further, $\alpha t_{T_P} \lambda_i \bar p_i$ denotes the \textit{revenue} paid by the secondary user $S_i$ to use the primary user's transmitter and is a function of the fixed (per unit) price $\alpha$, the service time fraction $t_{T_P,2}$ and the user's own `mean transmission rate' $\lambda_i \bar p_i$.     
\end{problem}

Now that we have formulated both the primary and secondary users' optimization problems, we note again that the coupling between all the users due to the shared ES makes the underlying problem a non-cooperative game. Thus, for the same reasons as discussed earlier, we will leverage the paradigm of MM-MFGs to compute tractable equilibrium policies for solving the $(N+1)$-user game problem. In this regard, we first need to characterize the average AoIs of both the primary and the secondary users which we do in the next section. Also, henceforth, we refer to $(p_i,\mu_{2i})$ as the policy of secondary device $D_{S_i}, \forall i$, and $(p_P,\mu_{1P},\mu_{2P})$ as the policy of primary device $D_{P}$.

\section{Average AoI Computation for MEC with Priority-Based Access}
In this section, we will compute the average AoIs of the primary as well as the secondary users.

\subsection{Average AoI Analysis for the Primary User}
Let us begin by considering the perspective of the primary user as shown in Fig. \ref{Fig:N_user_major_minor_single_major}, where we take the exogenous arrival process to be distributed as $\text{Poi}(\lambda_s)$ noting that $\lambda_s := \sum_{j=1}^N\lambda_j \bar p_j$, and all the incoming processes are Poisson distributed.

Next, we note that due to the priority access given to the primary device, on one hand, it can preempt the packet of any secondary device, and on the other hand, if a packet of the primary user is being served, the secondary user's task packet is dropped altogether. Thus, in essence, the secondary user does not have any effect on the average AoI of the primary user. Hence, for the purpose of computation, we can ignore the exogenous inputs due to the secondary users (i.e., set $\lambda_s = 0$ without loss of generality). Consequently, we can reduce the dimensionality of the FS-MC state space, which is elaborated upon as follows.

\textit{1. State Space:} The state space $\opS$ is comprised of 5 states which keep track of the server holding the freshest, the second freshest, and the oldest packets and/or is empty. Detailed descriptions are provided in Table \ref{table:states_major2}. We employ fake updates in both $L_P$ and the ES with the AoI of the fake packet equal to the AoI of the departing packet.

\begin{table}[t]
    \centering
    \begin{tabular}{ |c|c|c|c| } 
    \hline
    state & server 1 ($T_P$) & server 2 $(L_P)$ & server 3 (ES) \\
    \hline
    $s_1$ & freshest & $2^{nd}$ freshest & oldest \\ \hline
    $s_2$ & freshest & oldest & $2^{nd}$ freshest \\ \hline
    $s_3$ & $2^{nd}$ freshest & freshest & oldest \\ \hline
    $s_4$ & \emph{no packet} & freshest & $2^{nd}$ freshest  \\ \hline
    $s_5$ & \emph{no packet} & $2^{nd}$freshest & freshest \\
    \hline
    \end{tabular}
    \caption{\small{State dictionary for the FS-MC of the primary user.}}
    \label{table:states_major2}
\end{table}
    
\begin{table}[t]
    \centering
    \small
    \begin{tabular}{|c|c|c|c|c|} 
    \hline
    $s$ & ${\tt q}$ & $s'$ & $x' = xA_s$ & $v_s A_s$\\
    \hline
    \multirow{4}{2em}{$s_1$} & $\lambda_P p_P$ & $s_3$ & $[x_0 ~x_1 ~0 ~x_3]$ & $[\bar{v}_{10}~\bar{v}_{11}~0~\bar{v}_{13}]$\\
    & $\lambda_P\Bar{p}_P$ & $s_1$ & $[x_0 ~0 ~x_2 ~x_3]$ & $[\bar{v}_{10}~0~\bar{v}_{12}~\bar{v}_{13}]$ \\
    & $\mu_{1P}$ & $s_5$ & $[x_0 ~0 ~x_2 ~x_1]$ & $[\bar{v}_{10}~0~\bar{v}_{12}~\bar{v}_{11}]$ \\
    & $\mu_{2P}$ & $s_1$ & $[x_2 ~x_1 ~x_2 ~x_2]$ & $[\bar{v}_{12}~\bar{v}_{11}~\bar{v}_{12}~\bar{v}_{12}]$ \\
    & $\mu_3$ & $s_1$ & $[x_3 ~x_1 ~x_2 ~x_3]$ & $[\bar{v}_{13}~\bar{v}_{11}~\bar{v}_{12}~\bar{v}_{13}]$ \\ \hline
    \multirow{4}{2em}{$s_2$} & $\lambda_P p_P$ & $s_3$ & $[x_0 ~x_1 ~0 ~x_3]$ & $[\bar{v}_{20}~\bar{v}_{21}~0~\bar{v}_{23}]$ \\
    & $\lambda_P \Bar{p}_P$ & $s_2$ & $[x_0 ~0 ~x_2 ~x_3]$ & $[\bar{v}_{20}~0~\bar{v}_{22}~\bar{v}_{23}]$ \\
    & $\mu_{1P}$ & $s_5$ & $[x_0 ~0 ~x_2 ~x_1]$ & $[\bar{v}_{20}~0~\bar{v}_{22}~\bar{v}_{21}]$ \\
    & $\mu_{2P}$ & $s_2$ & $[x_2 ~x_1 ~x_2 ~x_3]$ & $[\bar{v}_{22}~\bar{v}_{21}~\bar{v}_{22}~\bar{v}_{23}]$ \\
    & $\mu_3$ & $s_2$ & $[x_3 ~x_1 ~x_3 ~x_3]$ & $[\bar{v}_{23}~\bar{v}_{21}~\bar{v}_{23}~\bar{v}_{23}]$ \\ \hline
    \multirow{4}{2em}{$s_3$} & $\lambda_P p_P$ & $s_3$ & $[x_0 ~x_1 ~0 ~x_3]$ & $[\bar{v}_{30}~\bar{v}_{31}~0~\bar{v}_{33}]$ \\
    & $\lambda_P \Bar{p}_P$ & $s_1$ & $[x_0 ~0 ~x_2 ~x_3]$ & $[\bar{v}_{30}~0~\bar{v}_{32}~\bar{v}_{33}]$ \\
    & $\mu_{1P}$ & $s_4$ & $[x_0 ~0 ~x_2 ~x_1]$ & $[\bar{v}_{30}~0~\bar{v}_{32}~\bar{v}_{31}]$ \\
    & $\mu_{2P}$ & $s_3$ & $[x_2 ~x_2 ~x_2 ~x_2]$ & $[\bar{v}_{32}~\bar{v}_{32}~\bar{v}_{32}~\bar{v}_{32}]$ \\
    & $\mu_3$ & $s_3$ & $[x_3 ~x_1 ~x_2 ~x_3]$ & $[\bar{v}_{33}~\bar{v}_{31}~\bar{v}_{32}~\bar{v}_{33}]$ \\ \hline
    \multirow{4}{2em}{$s_4$} & $\lambda_P p_P$ & $s_4$ & $[x_0 ~0 ~0 ~x_3]$ & $[\bar{v}_{40}~0~0~\bar{v}_{43}]$ \\
    & $\lambda_P\Bar{p}_P$ & $s_1$ & $[x_0 ~0 ~x_2 ~x_3]$ & $[\bar{v}_{40}~0~\bar{v}_{42}~\bar{v}_{43}]$ \\
    & $\mu_{2P}$ & $s_4$ & $[x_2 ~0 ~x_2 ~x_2]$ & $[\bar{v}_{42}~0~\bar{v}_{42}~\bar{v}_{42}]$ \\
    & $\mu_3$ & $s_4$& $[x_3 ~0 ~x_2 ~x_3]$ & $[\bar{v}_{43}~0~\bar{v}_{42}~\bar{v}_{43}]$ \\ \hline
    \multirow{4}{2em}{$s_5$} & $\lambda_P p_P$ & $s_4$ & $[x_0 ~0 ~0 ~x_3]$ & $[\bar{v}_{50}~0~0~\bar{v}_{53}]$ \\
    & $\lambda_P \Bar{p}_P$ & $s_2$ & $[x_0 ~0 ~x_2 ~x_3]$ & $[\bar{v}_{50}~0~\bar{v}_{52}~\bar{v}_{53}]$ \\
    & $\mu_{2P}$ & $s_5$ & $[x_2 ~0 ~x_2 ~x_3]$ & $[\bar{v}_{52}~0~\bar{v}_{52}~\bar{v}_{53}]$ \\
    & $\mu_3$ & $s_5$ & $[x_3 ~0 ~x_3 ~x_3]$ & $[\bar{v}_{53}~0~\bar{v}_{53}~\bar{v}_{53}]$ \\ \hline
    \end{tabular}
    \caption{\small{State transitions of the FS-MC and associated AoI jumps for the primary user.}}
    \label{table:transitions_major2}
\end{table}

\textit{2. Transition Functions:} Next, in Table \ref{table:transitions_major2}, we list the possible transitions in the FS-MC and the corresponding AoI vector $x'(t):= [x'_0(t) ~x'_1(t)~x'_2(t)~x'_3(t)]$. Let us take ${\tt u}_s := [1~1~1~1]$, $\hat {\tt u}_s := [1~0~1~1]$, $a:= \lambda_P + \mu_{1P} + \mu_{2P} + \mu_3$ and $\hat a := a - \mu_{1P}$. Then, the steady state vector $\Bar{\varrho}$ satisfies \eqref{prob_sum} and the following set of equations \eqref{eqn_P_a}-\eqref{eqn_P_e}.
\begin{subequations}\label{steady_state_prob_major2}
    \begin{align}\nonumber\\[-2.1em]
        a_P \Bar{\varrho}_1 & = (\lambda_P\Bar{p}_P + \mu_{2P} + \mu_3)\Bar{\varrho}_1 + \lambda_P\Bar{p}_P (\Bar{\varrho}_3 + \Bar{\varrho}_4), \label{eqn_P_a}\\ 
        a_P \Bar{\varrho}_2 & = (\lambda_P\Bar{p}_P + \mu_{2P} + \mu_3)\Bar{\varrho}_2 + \lambda_P\Bar{p}_P \Bar{\varrho}_5, \\ 
        a_P \Bar{\varrho}_3 & = (\lambda_P p_P + \mu_{2P} + \mu_3)\Bar{\varrho}_3 + \lambda_P p_P (\Bar{\varrho}_1 + \Bar{\varrho}_2), \\ 
        \hat a_P \Bar{\varrho}_4 & = (\lambda_P p_P + \mu_{2P} + \mu_3)\Bar{\varrho}_4 \!+\! \mu_{1P} \Bar{\varrho}_3 \!+\! \lambda_P p_P \Bar{\varrho}_5,  \\ 
        \hat a_P \Bar{\varrho}_5 & = (\mu_{2P} + \mu_3)\Bar{\varrho}_5 + \mu_{1P} (\Bar{\varrho}_1 + \Bar{\varrho}_2). \label{eqn_P_e}
    \end{align}
\end{subequations}

\begin{figure}[t]
    \centerline{\includegraphics[width=\columnwidth]{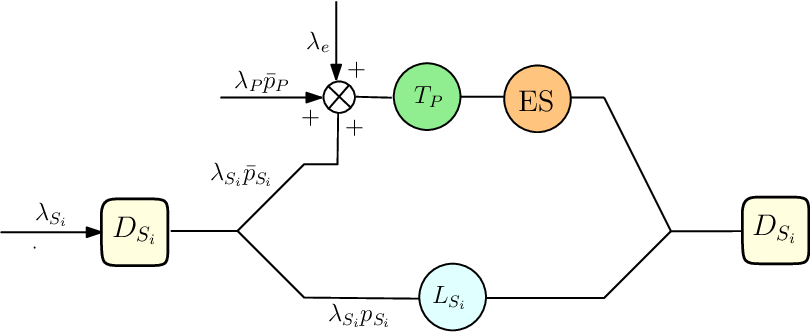}}
	\caption{\small{Task flow for secondary device $S_i$ with exogeneous incoming $\lambda_e$ and $\lambda_P \bar p_P$.}}
	\label{Fig:N_user_minor}
\end{figure}

\begin{figure*}[h]
\begin{subequations}\label{eqn_v_s_major2}
    \begin{align}
        a_P\Bar{v}_1 & \!= \!{\tt u}_s \Bar{\varrho}_1 \!+\! \lambda_P \Bar{p}_P[\bar{v}_{10}~0~\bar{v}_{12}~\bar{v}_{13}] +\mu_{2P} [\bar{v}_{12}~\bar{v}_{11}~\bar{v}_{12}~\bar{v}_{12}] +\lambda_P \Bar{p}_P[\bar{v}_{30}~0~\bar{v}_{32}~\bar{v}_{33}]   + \mu_3 [\bar{v}_{13}~\bar{v}_{11}~\bar{v}_{12}~\bar{v}_{13}] \nonumber \\
        & + \lambda_P \Bar{p}_P [\bar{v}_{40}~0~\bar{v}_{42}~\bar{v}_{43}] \\
        a_P\Bar{v}_2 & = {\tt u}_s \Bar{\varrho}_2 + \lambda_P \Bar{p}_P[\bar{v}_{20}~0~\bar{v}_{22}~\bar{v}_{23}] + \mu_{2P} [\bar{v}_{22}~\bar{v}_{21}~\bar{v}_{22}~\bar{v}_{23}] + \mu_3 [\bar{v}_{23}~\bar{v}_{21}~\bar{v}_{23}~\bar{v}_{23}] + \lambda_P \Bar{p}_P [\bar{v}_{50}~0~\bar{v}_{52}~\bar{v}_{53}] \\
        a_P\Bar{v}_3 & \!= \!{\tt u}_s \Bar{\varrho}_3 + \lambda_P p_P [\bar{v}_{30}~\bar{v}_{31}~0~\bar{v}_{33}] \!+ \!\mu_{2P} [\bar{v}_{32}~\bar{v}_{32}~\bar{v}_{32}~\bar{v}_{32}] + \mu_3 [\bar{v}_{33}~\bar{v}_{31}~\bar{v}_{32}~\bar{v}_{33}]  + \lambda_P p_P [\bar{v}_{10}~\bar{v}_{11}~0~\bar{v}_{13}]  \nonumber \\
        & + \lambda_P p_P [\bar{v}_{20}~\bar{v}_{21}~0~\bar{v}_{23}] \\
        \hat{a}_P\Bar{v}_4 & = \hat \opu_s \Bar{\varrho}_4 + \lambda_P p_P [\bar{v}_{40}~0~0~\bar{v}_{43}] \!+\! \mu_{2P} [\bar{v}_{42}~0~\bar{v}_{42}~\bar{v}_{42}] \!+\! \mu_3 [\bar{v}_{43}~0~\bar{v}_{42}~\bar{v}_{43}] \!+\! \lambda_P p_P [\bar{v}_{50}~0~0~\bar{v}_{53}]\! +\! \mu_{1P}[\bar{v}_{30}~0~\bar{v}_{32}~\bar{v}_{31}] \\
        \hat{a}_P\Bar{v}_5 & = \hat \opu_s \Bar{\varrho}_5 + \mu_{2P} [\bar{v}_{52}~0~\bar{v}_{52}~\bar{v}_{53}] + \mu_3 [\bar{v}_{53}~0~\bar{v}_{53}~\bar{v}_{53}] + \mu_{1P} [\bar{v}_{10}~0~\bar{v}_{12}~\bar{v}_{11}] + \mu_{1P} [\bar{v}_{20}~0~\bar{v}_{22}~\bar{v}_{21}]
    \end{align}
    \hrule
\end{subequations}
\end{figure*}

\begin{figure*}[h!]
\begin{subequations}
    \begin{align}\label{Avg_AOI_Primary}\nonumber\\[-3.7em]
        {\Delta}_P(p_P,\mu_{1P},\mu_{2P})  =& \frac{\mu_{1P} + \mu_{2P} + \mu_{3}}{(\mu_{1P} + \mu_{2P})(\mu_{2P} + \mu_{3})} + \frac{\mu_{1P}\mu_{3}(\lambda_P + \mu_{2P})}{\lambda_P (\mu_{1P} + \mu_{2P})(\mu_{2P} + \mu_{3}) (\lambda_P \times \bar p_P + \mu_{2P})} \nonumber \\
        & - \frac{\mu_{2P}\mu_{3}(\lambda_P + \mu_{1P})}{\lambda_P (\mu_{1P} + \mu_{2P})(\mu_{1P} - \mu_{3})(\mu_{1P} + \lambda_P \times p_P)} + \frac{\mu_{1P} \mu_{2P}(\lambda_P + \mu_{3})}{\lambda_P (\mu_{1P} - \mu_{3})(\mu_{2P} + \mu_{3})(\mu_{3} + \lambda_P \times p_P)}
    \end{align}
    \end{subequations}
    \hrule
\end{figure*}

Using the above set of linear equations in \eqref{steady_state_prob_major2}, one can compute the distribution $\bar \varrho$ by combining with \eqref{prob_sum}. To compute the average AoI ${\Delta}_P(p_P,\mu_{1P},\mu_{2P})$, it then remains to compute the steady state vector $\bar v$ in Theorem \ref{Avg_Age_thm} and then apply the formula for $\Delta$ in the theorem. To compute $\bar v$, we use \eqref{cond_prob_eqn} in Theorem \ref{Avg_Age_thm} to write down the set of linear equations satisfied by its components in \eqref{eqn_v_s_major2}. Consequently, the average AoI can be computed by first computing $\bar \varrho$ using \eqref{eqn_P_a}-\eqref{eqn_P_e}, substituting the same in \eqref{eqn_v_s_major2}, and subsequently solving the latter. We can then summarize the above result in the following theorem.
    
\begin{theorem}\label{thm:primary_user}
    Suppose that the arrival rate at the primary device is distributed as $\text{Poi}(\lambda_P)$ and the service rates as $\text{exp}(\mu_{1P})$ and $\text{exp}(\mu_{2P})$ at the transmitter $T_P$ and the local processor $L_P$, respectively. Let the service rate of the ES be distributed as $\text{exp}(\mu_3)$. Then, the average AoI of the primary user ${\Delta}_{P}(p_P,\mu_{1P},\mu_{2P})$ is given in \eqref{Avg_AOI_Primary} and is obtained by solving the sets of equations in \eqref{steady_state_prob_major2} and \eqref{eqn_v_s_major2}. 
\end{theorem}

The AoI computation for the primary user is now complete. Next, we proceed toward computing the average AoI for the secondary users.

\subsection{Average AoI Analysis for the Secondary User}
Let us now consider the perspective of a secondary user $S_i$ as shown in Fig. \ref{Fig:N_user_minor} with two types of exogeneous incomings, (1) a Poisson process with parameter $\lambda_P \bar p_P$ due to the primary user, and (2) a Poisson process with parameter $\lambda_e := \sum_{j \in [N], j \ne i} \lambda_j \bar p_j$ due to the other secondary users. The state space and the transition probabilities of the FS-MC for this system in Fig. \ref{Fig:N_user_minor} are given as follows.

\textit{1. State Space:} The state space $\opS$ is comprised of 10 states which keep track of the server holding the freshest, the second freshest, and the oldest packets of $S_i, \forall i$ and whether a server holds a packet from the primary user or not. We will refer to the packet of the primary user as `class P' packet. Detailed descriptions are provided in Table \ref{table:states_minor}. Furthermore, we run fake updates in all the servers where the fake packet is a low priority packet with its current AoI being the AoI of the departing packet.

\begin{table}[t]
    \centering
    \begin{tabular}{ |c|c|c|c| } 
    \hline
    state & server 1 ($T_{P}$) & server 2 $(L_{S_i})$ & server 3 (ES) \\
    \hline
    $s_1$ & freshest & $2^{nd}$ freshest & oldest \\ \hline
    $s_2$ & freshest & oldest & $2^{nd}$ freshest \\ \hline
    $s_3$ & freshest & $2^{nd}$ freshest & class P \\ \hline
    $s_4$ & $2^{nd}$ freshest & freshest & oldest  \\ \hline
    $s_5$ & class P & freshest & $2^{nd}$ freshest \\ \hline
    $s_6$ & $2^{nd}$ freshest & freshest & class P \\ \hline
    $s_7$ & class P & freshest & class P \\ \hline
    $s_8$ & class P & $2^{nd}$ freshest & freshest \\ \hline
    $s_9$ & oldest & freshest & $2^{nd}$ freshest \\ \hline
    $s_{10}$ & oldest & $2^{nd}$ freshest & freshest \\ 
    \hline
    \end{tabular}
    \caption{\small{\!\!State dictionary for the FS-MC of the secondary user $\!S_i$.}}
    \label{table:states_minor}
\end{table}

\textit{2. Transition functions:} Next, in Tables \ref{table:transitions_minor_T1}, we list the possible transitions in the FS-MC and the corresponding AoI vector $x'(t):= [x'_0(t) ~x'_1(t)~x'_2(t)~x'_3(t)]$, where $x_0',x_1',x_2'$, and $x_3'$ denote the AoIs of the packets at the device $D_{S_i}, T_P, L_{S_i}$, and the ES, respectively. Henceforth, we will suppress the index $i$ for brevity. Similar to the earlier analysis, let us take ${\tt u}_s := [1~1~1~1]$ and $a\!:=\! \lambda + \lambda_e \! +\!  \mu_1 \! +\!  \mu_2\!  +\!  \mu_3$. Then, the steady state vector $\Bar{\pi}$ for the secondary user $S_i$ satisfies \eqref{prob_sum} and the following set of equations:
\begin{subequations}\label{steady_state_prob_minor}
    \begin{align}
        a \Bar{\pi}_1  =& (\lambda\Bar{p} + \mu_{2} + \mu_3)\Bar{\pi}_1 + \lambda\Bar{p} (\Bar{\pi}_4 + \Bar{\pi}_9) + \mu_3 \bar \pi_3, \\ 
        a \Bar{\pi}_2  =& (\lambda\Bar{p} + \mu_{2} + \mu_3)\Bar{\pi}_2 + \lambda\Bar{p} \Bar{\pi}_{10}, \\ 
        a \Bar{\pi}_3  =& (\lambda \bar p + \mu_{2} )\Bar{\pi}_3 + \lambda \bar p \Bar{\pi}_6, \\ 
         a \Bar{\pi}_4  =& (\lambda p + \mu_{2} + \mu_3)\Bar{\pi}_4 \!+\! \mu_3\Bar{\pi}_6 \!+\! \lambda p (\Bar{\pi}_1 + \bar \pi_2),  \\ 
         a \Bar{\pi}_5  =& (a \!-\! \mu_1)\Bar{\pi}_5 \!+\! \lambda_P (\Bar{\pi}_1 + \Bar{\pi}_4 + \Bar{\pi}_9) + \mu_3 \bar \pi_7 + \lambda p \pi_8, \\
         a \pi_6   =& (\lambda p + \lambda_e + \mu_1 + \mu_2) \bar \pi_6 + (\lambda p + \lambda_e + \mu_1) \bar \pi_3 \nonumber \\
         & + \mu_1(\bar \pi_5 + \bar \pi_7 + \bar \pi_8), \\
         a \bar \pi_7  =&  (a - \mu_1 - \mu_3)\bar \pi_7 + \lambda_P (\bar \pi_3 + \bar \pi_6), \\
         a \bar \pi_8  = &(a - \lambda p -\mu_1) \bar \pi_8 + \lambda_P (\bar \pi_2 + \bar \pi_{10}), \\
         a \bar \pi_9 =& (a- \lambda \bar p - \lambda_P) \bar \pi_9 + \lambda_e (\bar \pi_1 + \bar \pi_4) + \mu_1 \bar \pi_4 \nonumber \\
         & + (\lambda p + \mu_1) \bar \pi_{10}, \\
         a \bar \pi_{10}  = &(\lambda_e + \mu_2 + \mu_3) \bar \pi_{10} + \mu_1 (\bar \pi_1 + \bar \pi_2) + \lambda_e \bar \pi_2.
    \end{align}
\end{subequations}
Subsequently, using \eqref{cond_prob_eqn}, we can use the solution to \eqref{steady_state_prob_minor} to write down the equations satisfied by the steady-state conditional distribution vector for the secondary user $S_i$ as in \eqref{eqn_v_s_minor}. Then, we solve this linear system of equations to obtain the average AoI $\Delta^{(N)}_i\!(\bm{p_{ps}}, \bm{\mu_{2,ps}}, \bm{\mu_{1P}})$ which we summarize in the next theorem by resuming the use of notation $i$ for indexing.

\begin{table*}[t]
    \small
    \parbox{.49\linewidth}{
    \centering
    \begin{tabular}{|c|c|c|c|c|} 
    \hline
    $s$ & $\opq$ & $s'$ & $x' = xA_s$ & $v_s A_s$\\
    \hline
    \multirow{4}{2em}{$s_1$} & $\lambda p$ & $s_4$ & $[x_0 ~x_1 ~0 ~x_3]$ & $[\bar{v}_{10}~\bar{v}_{11}~0~\bar{v}_{13}]$\\
    & $\lambda\Bar{p}$ & $s_1$ & $[x_0 ~0 ~x_2 ~x_3]$ & $[\bar{v}_{10}~0~\bar{v}_{12}~\bar{v}_{13}]$ \\
    & $\lambda_e$ & $s_9$ & $[x_0 ~x_0 ~x_2 ~x_3]$ & $[\bar{v}_{10}~\bar{v}_{10}~\bar{v}_{12}~\bar{v}_{13}]$ \\
    & $\lambda_P$ & $s_5$ & $[x_0 ~x_0 ~x_2 ~x_3]$ & $[\bar{v}_{10}~\bar{v}_{10}~\bar{v}_{12}~\bar{v}_{13}]$ \\
    & $\mu_1$ & $s_{10}$ & $[x_0 ~x_1 ~x_2 ~x_1]$ & $[\bar{v}_{10}~\bar{v}_{11}~\bar{v}_{12}~\bar{v}_{11}]$ \\
    & $\mu_2$ & $s_1$ & $[x_2 ~x_1 ~x_2 ~x_2]$ & $[\bar{v}_{12}~\bar{v}_{11}~\bar{v}_{12}~\bar{v}_{12}]$ \\
    & $\mu_3$ & $s_1$ & $[x_3 ~x_1 ~x_2 ~x_3]$ & $[\bar{v}_{13}~\bar{v}_{11}~\bar{v}_{12}~\bar{v}_{13}]$ \\ \hline
    \multirow{4}{2em}{$s_2$} & $\lambda p$ & $s_4$ & $[x_0 ~x_1 ~0 ~x_3]$ & $[\bar{v}_{20}~\bar{v}_{21}~0~\bar{v}_{23}]$ \\
    & $\lambda \Bar{p}$ & $s_2$ & $[x_0 ~0 ~x_2 ~x_3]$ & $[\bar{v}_{20}~0~\bar{v}_{22}~\bar{v}_{23}]$ \\
    & $\lambda_e$ & $s_{10}$ & $[x_0 ~x_0 ~x_2 ~x_3]$ & $[\bar{v}_{20}~\bar{v}_{20}~\bar{v}_{22}~\bar{v}_{23}]$ \\
    & $\lambda_P$ & $s_8$ & $[x_0 ~x_0 ~x_2 ~x_3]$ & $[\bar{v}_{20}~\bar{v}_{20}~\bar{v}_{22}~\bar{v}_{23}]$ \\
    & $\mu_1$ & $s_{10}$ & $[x_0 ~x_1 ~x_2 ~x_1]$ & $[\bar{v}_{20}~\bar{v}_{21}~\bar{v}_{22}~\bar{v}_{21}]$ \\
    & $\mu_2$ & $s_2$ & $[x_2 ~x_1 ~x_2 ~x_3]$ & $[\bar{v}_{22}~\bar{v}_{21}~\bar{v}_{22}~\bar{v}_{23}]$ \\
    & $\mu_3$ & $s_2$ & $[x_3 ~x_1 ~x_3 ~x_3]$ & $[\bar{v}_{23}~\bar{v}_{21}~\bar{v}_{23}~\bar{v}_{23}]$ \\ \hline
    \multirow{4}{2em}{$s_3$} & $\lambda p$ & $s_6$ & $[x_0 ~x_1 ~0 ~x_3]$ & $[\bar{v}_{30}~\bar{v}_{31}~0~\bar{v}_{33}]$ \\
    & $\lambda \Bar{p}$ & $s_3$ & $[x_0 ~0 ~x_2 ~x_3]$ & $[\bar{v}_{30}~0~\bar{v}_{32}~\bar{v}_{33}]$ \\
    & $\lambda_e$ & $s_6$ & $[x_0 ~x_0 ~x_2 ~x_3]$ & $[\bar{v}_{30}~\bar{v}_{30}~\bar{v}_{32}~\bar{v}_{33}]$ \\
    & $\lambda_P$ & $s_7$ & $[x_0 ~x_0 ~x_2 ~x_3]$ & $[\bar{v}_{30}~\bar{v}_{30}~\bar{v}_{32}~\bar{v}_{33}]$ \\
    & $\mu_1$ & $s_6$ & $[x_0 ~x_0 ~x_2 ~x_3]$ & $[\bar{v}_{30}~\bar{v}_{30}~\bar{v}_{32}~\bar{v}_{33}]$ \\
    & $\mu_2$ & $s_3$ & $[x_2 ~x_1 ~x_2 ~x_2]$ & $[\bar{v}_{32}~\bar{v}_{31}~\bar{v}_{32}~\bar{v}_{32}]$ \\
    & $\mu_3$& $s_1$  & $[x_3 ~x_1 ~x_2 ~x_3]$ & $[\bar{v}_{33}~\bar{v}_{31}~\bar{v}_{32}~\bar{v}_{33}]$ \\ \hline
    \multirow{4}{2em}{$s_4$} & $\lambda p$ & $s_4$ & $[x_0 ~x_1 ~0 ~x_3]$ & $[\bar{v}_{40}~\bar{v}_{41}~0~\bar{v}_{43}]$ \\
    & $\lambda\Bar{p}$ & $s_1$ & $[x_0 ~0 ~x_2 ~x_3]$ & $[\bar{v}_{40}~0~\bar{v}_{42}~\bar{v}_{43}]$ \\
    & $\lambda_e$ & $s_9$ & $[x_0 ~x_0 ~x_2 ~x_3]$ & $[\bar{v}_{40}~\bar{v}_{40}~\bar{v}_{42}~\bar{v}_{43}]$ \\
    & $\lambda_P$ & $s_5$ & $[x_0 ~x_0 ~x_2 ~x_3]$ & $[\bar{v}_{40}~\bar{v}_{40}~\bar{v}_{42}~\bar{v}_{43}]$ \\
    & $\mu_1$ & $s_9$ & $[x_0 ~x_1 ~x_2 ~x_1]$ & $[\bar{v}_{40}~\bar{v}_{41}~\bar{v}_{42}~\bar{v}_{41}]$ \\
    & $\mu_2$ & $s_4$ & $[x_2 ~x_2 ~x_2 ~x_2]$ & $[\bar{v}_{42}~\bar{v}_{42}~\bar{v}_{42}~\bar{v}_{42}]$ \\
    & $\mu_3$ & $s_4$ & $[x_3 ~x_1 ~x_2 ~x_3]$ & $[\bar{v}_{43}~\bar{v}_{41}~\bar{v}_{42}~\bar{v}_{43}]$ \\ \hline
    \multirow{4}{2em}{$s_5$} & $\lambda p$ & $s_5$ & $[x_0 ~x_1 ~0 ~x_3]$ & $[\bar{v}_{50}~\bar{v}_{51}~0~\bar{v}_{53}]$ \\
    & $\lambda \Bar{p}$ & $s_5$ & $[x_0 ~x_1 ~x_2 ~x_3]$ & $[\bar{v}_{50}~\bar{v}_{51}~\bar{v}_{52}~\bar{v}_{53}]$ \\
    & $\lambda_e$ & $s_5$ & $[x_0 ~x_1 ~x_2 ~x_3]$ & $[\bar{v}_{50}~\bar{v}_{51}~\bar{v}_{52}~\bar{v}_{53}]$ \\
    & $\lambda_P$ & $s_5$ & $[x_0 ~x_0 ~x_2 ~x_3]$ & $[\bar{v}_{50}~\bar{v}_{50}~\bar{v}_{52}~\bar{v}_{53}]$ \\
    & $\mu_1$ & $s_6$ & $[x_0 ~x_1 ~x_2 ~x_1]$ & $[\bar{v}_{50}~\bar{v}_{51}~\bar{v}_{52}~\bar{v}_{51}]$ \\
    & $\mu_2$ & $s_5$ & $[x_2 ~x_2 ~x_2 ~x_2]$ & $[\bar{v}_{52}~\bar{v}_{52}~\bar{v}_{52}~\bar{v}_{52}]$ \\
    & $\mu_3$ & $s_5$ & $[x_3 ~x_3 ~x_2 ~x_3]$ & $[\bar{v}_{53}~\bar{v}_{53}~\bar{v}_{52}~\bar{v}_{53}]$ \\ \hline
    \end{tabular}}
    \parbox{.49\linewidth}{
    \centering
    \begin{tabular}{|c|c|c|c|c|} 
    \hline
    $s$ & $\opq$ & $s'$ & $x' = xA_s$ & $v_s A_s$\\
    \hline
    \multirow{4}{2em}{$s_6$} & $\lambda p$ & $s_6$ & $[x_0 ~x_1 ~0 ~x_3]$ & $[\bar{v}_{60}~\bar{v}_{61}~0~\bar{v}_{63}]$ \\
    & $\lambda \Bar{p}$ & $s_3$ & $[x_0 ~0 ~x_2 ~x_3]$ & $[\bar{v}_{60}~0~\bar{v}_{62}~\bar{v}_{63}]$ \\
    & $\lambda_e$ & $s_6$ & $[x_0 ~x_0 ~x_2 ~x_3]$ & $[\bar{v}_{60}~\bar{v}_{60}~\bar{v}_{62}~\bar{v}_{63}]$ \\
    & $\lambda_P$ & $s_7$ & $[x_0 ~x_0 ~x_2 ~x_3]$ & $[\bar{v}_{60}~\bar{v}_{60}~\bar{v}_{62}~\bar{v}_{63}]$ \\
    & $\mu_1$ & $s_6$ & $[x_0 ~x_0 ~x_2 ~x_3]$ & $[\bar{v}_{60}~\bar{v}_{60}~\bar{v}_{62}~\bar{v}_{63}]$ \\
    & $\mu_2$ & $s_6$ & $[x_2 ~x_2 ~x_2 ~x_2]$ & $[\bar{v}_{62}~\bar{v}_{62}~\bar{v}_{62}~\bar{v}_{62}]$ \\
    & $\mu_3$ & $s_4$ & $[x_3 ~x_1 ~x_2 ~x_3]$ & $[\bar{v}_{63}~\bar{v}_{61}~\bar{v}_{62}~\bar{v}_{63}]$ \\ \hline
    \multirow{4}{2em}{$s_7$} & $\lambda p$ & $s_7$ & $[x_0 ~x_1 ~0 ~x_3]$ & $[\bar{v}_{70}~\bar{v}_{71}~0~\bar{v}_{73}]$ \\
    & $\lambda \Bar{p}$ & $s_7$ & $[x_0 ~x_1 ~x_2 ~x_3]$ & $[\bar{v}_{70}~\bar{v}_{71}~\bar{v}_{72}~\bar{v}_{73}]$ \\
    & $\lambda_e$ & $s_7$ & $[x_0 ~x_1 ~x_2 ~x_3]$ & $[\bar{v}_{70}~\bar{v}_{71}~\bar{v}_{72}~\bar{v}_{73}]$ \\
    & $\lambda_P$ & $s_7$ & $[x_0 ~x_0 ~x_2 ~x_3]$ & $[\bar{v}_{70}~\bar{v}_{70}~\bar{v}_{72}~\bar{v}_{73}]$ \\
    & $\mu_1$ & $s_6$ & $[x_0 ~x_1 ~x_2 ~x_1]$ & $[\bar{v}_{70}~\bar{v}_{71}~\bar{v}_{72}~\bar{v}_{71}]$ \\
    & $\mu_2$ & $s_7$ & $[x_2 ~x_2 ~x_2 ~x_2]$ & $[\bar{v}_{72}~\bar{v}_{72}~\bar{v}_{72}~\bar{v}_{72}]$ \\
    & $\mu_3$ & $s_5$ & $[x_3 ~x_3 ~x_2 ~x_3]$ & $[\bar{v}_{73}~\bar{v}_{73}~\bar{v}_{72}~\bar{v}_{73}]$ \\ \hline
    \multirow{4}{2em}{$s_8$} & $\lambda p$ & $s_5$ & $[x_0 ~x_1 ~0 ~x_3]$ & $[\bar{v}_{80}~\bar{v}_{81}~0~\bar{v}_{83}]$ \\
    & $\lambda \Bar{p}$ & $s_8$ & $[x_0 ~x_1 ~x_2 ~x_3]$ & $[\bar{v}_{80}~\bar{v}_{81}~\bar{v}_{82}~\bar{v}_{83}]$ \\
    & $\lambda_e$ & $s_8$ & $[x_0 ~x_1 ~x_2 ~x_3]$ & $[\bar{v}_{80}~\bar{v}_{81}~\bar{v}_{82}~\bar{v}_{83}]$ \\
    & $\lambda_P$ & $s_8$ & $[x_0 ~x_0 ~x_2 ~x_3]$ & $[\bar{v}_{80}~\bar{v}_{80}~\bar{v}_{82}~\bar{v}_{83}]$ \\
    & $\mu_1$ & $s_6$ & $[x_0 ~x_1 ~x_2 ~x_1]$ & $[\bar{v}_{80}~\bar{v}_{81}~\bar{v}_{82}~\bar{v}_{81}]$ \\
    & $\mu_2$ & $s_8$ & $[x_2 ~x_2 ~x_2 ~x_3]$ & $[\bar{v}_{82}~\bar{v}_{82}~\bar{v}_{82}~\bar{v}_{83}]$ \\
    & $\mu_3$ & $s_8$ & $[x_3 ~x_3 ~x_3 ~x_3]$ & $[\bar{v}_{83}~\bar{v}_{83}~\bar{v}_{83}~\bar{v}_{83}]$ \\ \hline
    \multirow{4}{2em}{$s_9$} & $\lambda p$ & $s_9$ & $[x_0 ~x_1 ~0 ~x_3]$ & $[\bar{v}_{90}~\bar{v}_{91}~0~\bar{v}_{93}]$ \\
    & $\lambda \Bar{p}$ & $s_1$ & $[x_0 ~0 ~x_2 ~x_3]$ & $[\bar{v}_{90}~0~\bar{v}_{92}~\bar{v}_{93}]$ \\
    & $\lambda_e$ & $s_9$ & $[x_0 ~x_0 ~x_2 ~x_3]$ & $[\bar{v}_{90}~\bar{v}_{90}~\bar{v}_{92}~\bar{v}_{93}]$ \\
    & $\lambda_P$ & $s_5$ & $[x_0 ~x_0 ~x_2 ~x_3]$ & $[\bar{v}_{90}~\bar{v}_{90}~\bar{v}_{92}~\bar{v}_{93}]$ \\
    & $\mu_1$ & $s_9$ & $[x_0 ~x_1 ~x_2 ~x_1]$ & $[\bar{v}_{90}~\bar{v}_{91}~\bar{v}_{92}~\bar{v}_{91}]$ \\
    & $\mu_2$ & $s_9$ & $[x_2 ~x_2 ~x_2 ~x_2]$ & $[\bar{v}_{92}~\bar{v}_{92}~\bar{v}_{92}~\bar{v}_{92}]$ \\
    & $\mu_3$ & $s_9$ & $[x_3 ~x_3 ~x_2 ~x_3]$ & $[\bar{v}_{93}~\bar{v}_{93}~\bar{v}_{92}~\bar{v}_{93}]$ \\ \hline
    \multirow{4}{2em}{$s_{10}$} & $\lambda p$ & $s_9$ & $[x_0 ~x_1 ~0 ~x_3]$ & $[\bar{v}_{10,0}~\bar{v}_{10,1}~0~\bar{v}_{10,3}]$ \\
    & $\lambda \Bar{p}$ & $s_2$ & $[x_0 ~0 ~x_2 ~x_3]$ & $[\bar{v}_{10,0}~0~\bar{v}_{10,2}~\bar{v}_{10,3}]$ \\
    & $\lambda_e$ & $s_{10}$ & $[x_0 ~x_0 ~x_2 ~x_3]$ & $[\bar{v}_{10,0}~\bar{v}_{10,0}~\bar{v}_{10,2}~\bar{v}_{10,3}]$ \\
    & $\lambda_P$ & $s_8$ & $[x_0 ~x_0 ~x_2 ~x_3]$ & $[\bar{v}_{10,0}~\bar{v}_{10,0}~\bar{v}_{10,2}~\bar{v}_{10,3}]$ \\
    & $\mu_1$ & $s_9$ & $[x_0 ~x_1 ~x_2 ~x_1]$ & $[\bar{v}_{10,0}~\bar{v}_{10,1}~\bar{v}_{10,2}~\bar{v}_{10,1}]$ \\
    & $\mu_2$ & $s_{10}$ & $[x_2 ~x_2 ~x_2 ~x_3]$ & $[\bar{v}_{10,2}~\bar{v}_{10,2}~\bar{v}_{10,2}~\bar{v}_{10,3}]$ \\
    & $\mu_3$ & $s_{10}$ & $[x_3 ~x_3 ~x_3 ~x_3]$ & $[\bar{v}_{10,3}~\bar{v}_{10,3}~\bar{v}_{10,3}~\bar{v}_{10,3}]$ \\ \hline
    \end{tabular}}
    \caption{\small{State transitions of the FS-MC and associated AoI jumps for the secondary user.}}
    \label{table:transitions_minor_T1}
\end{table*}

\begin{figure*}[h]
\begin{subequations}\label{eqn_v_s_minor}
\begin{small}
    \begin{align}
        a\Bar{v}_1 & \!\!= \!{\tt u}_s \Bar{\pi}_1 \!\!+\!\! \lambda \Bar{p}[\bar{v}_{10}~0~\bar{v}_{12}~\bar{v}_{13}] \!+\! \mu_3 [\bar{v}_{13}~\bar{v}_{11}~\bar{v}_{12}~\bar{v}_{13}]\! +\! \mu_3[\bar{v}_{33}~\bar{v}_{31}~\bar{v}_{32}~\bar{v}_{33}] \!+\! \mu_2 [\bar{v}_{12}~\bar{v}_{11}~\bar{v}_{12}~\bar{v}_{12}] \!+\! \lambda \Bar{p} [\bar{v}_{40}~0~\bar{v}_{42}~\bar{v}_{43}]\!  \nonumber \\&~~+ \lambda \Bar{p} [\bar{v}_{90}~0~\bar{v}_{92}~\bar{v}_{93}] \\
        a\Bar{v}_2 & = {\tt u}_s \Bar{\pi}_2 + \lambda \Bar{p}[\bar{v}_{20}~0~\bar{v}_{22}~\bar{v}_{23}] + \mu_2 [\bar{v}_{22}~\bar{v}_{21}~\bar{v}_{22}~\bar{v}_{23}] + \mu_3 [\bar{v}_{23}~\bar{v}_{21}~\bar{v}_{23}~\bar{v}_{23}] + \lambda \Bar{p} [\bar{v}_{10,0}~0~\bar{v}_{10,2}~\bar{v}_{10,3}] \\
        a\Bar{v}_3 & \!= \!{\tt u}_s \Bar{\pi}_3 + \lambda \bar p [\bar{v}_{30}~0~\bar{v}_{32}~\bar{v}_{33}] \!+ \!\mu_2 [\bar{v}_{32}~\bar{v}_{31}~\bar{v}_{32}~\bar{v}_{32}] + \lambda \bar p [\bar{v}_{60}~0~\bar{v}_{62}~\bar{v}_{63}] \\
        {a}\Bar{v}_4 & = {\operatorname{u}}_s \Bar{\pi}_4 + \lambda p [\bar{v}_{40}~\bar{v}_{41}~0~\bar{v}_{43}] + \mu_2 [\bar{v}_{42}~\bar{v}_{42}~\bar{v}_{42}~\bar{v}_{42}] + \mu_3 [\bar{v}_{43}~\bar{v}_{41}~\bar{v}_{42}~\bar{v}_{43}] + \mu_3[\bar{v}_{63}~\bar{v}_{61}~\bar{v}_{62}~\bar{v}_{63}] + \lambda p [\bar{v}_{10}~\bar{v}_{11}~0~\bar{v}_{13}] \nonumber \\
        & ~~+ \lambda p [\bar{v}_{20}~\bar{v}_{21}~0~\bar{v}_{23}]\\
        {a}\Bar{v}_5 & = {\tt u}_s \Bar{\pi}_5 + \lambda p [\bar{v}_{50}~\bar{v}_{51}~0~\bar{v}_{53}] + \mu_2 [\bar{v}_{52}~\bar{v}_{52}~\bar{v}_{52}~\bar{v}_{52}] + \mu_3 [\bar{v}_{53}~\bar{v}_{53}~\bar{v}_{52}~\bar{v}_{53}] + \mu_3 [\bar{v}_{73}~\bar{v}_{73}~\bar{v}_{72}~\bar{v}_{73}] + \lambda p [\bar{v}_{80}~\bar{v}_{81}~0~\bar{v}_{83}] \nonumber \\
        & ~~ + (\lambda \bar p + \lambda_e) [\bar{v}_{50}~\bar{v}_{51}~\bar{v}_{52}~\bar{v}_{53}] + \lambda_P [\bar{v}_{10}~\bar{v}_{10}~\bar{v}_{12}~\bar{v}_{13}] + \lambda_P [\bar{v}_{40}~\bar{v}_{40}~\bar{v}_{42}~\bar{v}_{43}]+ \lambda_P [\bar{v}_{90}~\bar{v}_{90}~\bar{v}_{92}~\bar{v}_{93}]+ \lambda_P [\bar{v}_{50}~\bar{v}_{50}~\bar{v}_{52}~\bar{v}_{53}]\\
        {a}\Bar{v}_6 & = {\tt u}_s \Bar{\pi}_6 + \lambda p [\bar{v}_{60}~\bar{v}_{61}~0~\bar{v}_{63}] + \lambda p [\bar{v}_{30}~\bar{v}_{31}~0~\bar{v}_{33}] + \mu_2 [\bar{v}_{62}~\bar{v}_{62}~\bar{v}_{62}~\bar{v}_{62}] + \mu_1 [\bar{v}_{60}~\bar{v}_{60}~\bar{v}_{62}~\bar{v}_{63}] + \mu_1 [\bar{v}_{50}~\bar{v}_{51}~\bar{v}_{52}~\bar{v}_{51}] \nonumber \\
        & ~~ + \mu_1 [\bar{v}_{30}~\bar{v}_{30}~\bar{v}_{32}~\bar{v}_{33}] + \lambda_e [\bar{v}_{60}~\bar{v}_{60}~\bar{v}_{62}~\bar{v}_{63}] + \lambda_e [\bar{v}_{30}~\bar{v}_{30}~\bar{v}_{32}~\bar{v}_{33}] + \mu_1 [\bar{v}_{70}~\bar{v}_{71}~\bar{v}_{72}~\bar{v}_{71}] + \mu_1 [\bar{v}_{80}~\bar{v}_{81}~\bar{v}_{82}~\bar{v}_{81}]\\
        a\Bar{v}_7 & = {\tt u}_s \Bar{\pi}_7 + \lambda p [\bar{v}_{70}~\bar{v}_{71}~0~\bar{v}_{73}] + \lambda \bar p [\bar{v}_{70}~\bar{v}_{71}~\bar{v}_{72}~\bar{v}_{73}] + \lambda_e [\bar{v}_{70}~\bar{v}_{71}~\bar{v}_{72}~\bar{v}_{73}] + \mu_2 [\bar{v}_{72}~\bar{v}_{72}~\bar{v}_{72}~\bar{v}_{72}] + \lambda_P [\bar{v}_{70}~\bar{v}_{70}~\bar{v}_{72}~\bar{v}_{73}] \nonumber \\
        & ~~ + \lambda_P [\bar{v}_{30}~\bar{v}_{30}~\bar{v}_{32}~\bar{v}_{33}] + \lambda_P [\bar{v}_{60}~\bar{v}_{60}~\bar{v}_{62}~\bar{v}_{63}]  \\
        a\Bar{v}_8 & = {\tt u}_s \Bar{\pi}_8 + \lambda \bar{p} [\bar{v}_{80}~\bar{v}_{81}~\bar{v}_{82}~\bar{v}_{83}] + \lambda_e [\bar{v}_{80}~\bar{v}_{81}~\bar{v}_{82}~\bar{v}_{83}] + \mu_2 [\bar{v}_{82}~\bar{v}_{82}~\bar{v}_{82}~\bar{v}_{83}]+ \mu_3 [\bar{v}_{83}~\bar{v}_{83}~\bar{v}_{83}~\bar{v}_{83}] \nonumber \\
        & ~~ + \lambda_P [\bar{v}_{20}~\bar{v}_{20}~\bar{v}_{22}~\bar{v}_{23}] + \lambda_P [\bar{v}_{80}~\bar{v}_{80}~\bar{v}_{82}~\bar{v}_{83}]+ \lambda_P [\bar{v}_{10,0}~\bar{v}_{10,0}~\bar{v}_{10,2}~\bar{v}_{10,3}] \\
        a\Bar{v}_9 & = {\tt u}_s \Bar{\pi}_9 + \lambda p [\bar{v}_{90}~\bar{v}_{91}~0~\bar{v}_{93}] + \lambda_e [\bar{v}_{90}~\bar{v}_{90}~\bar{v}_{92}~\bar{v}_{93}] + \mu_2 [\bar{v}_{92}~\bar{v}_{92}~\bar{v}_{92}~\bar{v}_{92}]+ \mu_3 [\bar{v}_{93}~\bar{v}_{93}~\bar{v}_{92}~\bar{v}_{93}] + \mu_1 [\bar{v}_{40}~\bar{v}_{41}~\bar{v}_{42}~\bar{v}_{41}]\nonumber \\
        & ~~ + \mu_1 [\bar{v}_{90}~\bar{v}_{91}~\bar{v}_{92}~\bar{v}_{91}] + \lambda_e [\bar{v}_{10}~\bar{v}_{10}~\bar{v}_{12}~\bar{v}_{13}] + \lambda p [\bar{v}_{10,0}~\bar{v}_{10,1}~0~\bar{v}_{10,3}] + \mu_1 [\bar{v}_{10,0}~\bar{v}_{10,1}~\bar{v}_{10,2}~\bar{v}_{10,1}] + \lambda_e [\bar{v}_{40}~\bar{v}_{40}~\bar{v}_{42}~\bar{v}_{43}]\\
        a\Bar{v}_{10} & = {\tt u}_s \Bar{\pi}_{10} + \lambda_e [\bar{v}_{10,0}~\bar{v}_{10,0}~\bar{v}_{10,2}~\bar{v}_{10,3}] + \lambda_e [\bar{v}_{20}~\bar{v}_{20}~\bar{v}_{22}~\bar{v}_{23}] + \mu_2 [\bar{v}_{10,2}~\bar{v}_{10,2}~\bar{v}_{10,2}~\bar{v}_{10,3}] + \mu_3[\bar{v}_{10,3}~\bar{v}_{10,3}~\bar{v}_{10,3}~\bar{v}_{10,3}] \nonumber \\
        & ~~ + \mu_1 [\bar{v}_{10}~\bar{v}_{11}~\bar{v}_{12}~\bar{v}_{11}] + \mu_1 [\bar{v}_{20}~\bar{v}_{21}~\bar{v}_{22}~\bar{v}_{21}]
    \end{align}
    \end{small}
    \hrule
\end{subequations}
\end{figure*}

\begin{theorem}\label{thm:Secon_user}
    Suppose that the arrival rate at a secondary device $S_i$ is distributed as $\text{Poi}(\lambda_{S_i})$ and the service rate of its local processor as $\text{exp}(\mu_{2i})$. Further, let the service rates of the transmitter of the primary user and the ES be distributed as $\text{exp}(\mu_{1P})$ and $\text{exp}(\mu_3)$, respectively. Then, the average AoI ${\Delta}^{(N)}_{i}(\bm{p_{ps}},\bm{\mu_{2,ps}},\bm{\mu_{1P}})$ of $S_i$  exists for all $i$ and is obtained by solving the sets of equations in \eqref{steady_state_prob_minor} and \eqref{eqn_v_s_minor}.
\end{theorem}

Theorem \ref{thm:primary_user} and Theorem \ref{thm:Secon_user} completely characterize the average AoI of the primary and the secondary users, respectively, and hence, also Problems \ref{problem:primary_user} and \ref{problem:sec_user}. Due to long expressions of ${\Delta}^{(N)}_{i}(\bm{p_{ps}},\bm{\mu_{2,ps}},\bm{\mu_{1P}})$, we do not provide them here; however, later, we will provide an algorithm to compute the equilibrium policies without requiring their explicit forms. We are now ready to provide equilibrium policies for the MEC system with priority-based access.

\section{Major-Minor Mean-Field Analysis for the MEC System With Priority Access}
Our aim here is again to compute Nash equilibrium policies for the primary and the secondary users. However, for reasons of tractability discussed before, we appeal to the MM-MFG framework to compute completely \textit{distributed} approximate NE solutions using only the local information of the users.

To this end, let us define, using the same notations as in previous sections, the quantities $\rho^{(N)} := \frac{\lambda_e}{N \mu_{1P}}$ and $\rho := \lim_{N \rightarrow \infty}\rho^{(N)}$, where $\rho^{(N)}$ denotes the mean load on the primary device's transmitter $T_P$ and $\rho$ denotes the MF approximation as discussed earlier. Then, we have that for a large user MEC system, the exogenous arrival rates $\lambda_e$ and $\lambda_s$ can be approximated as
\begin{align}
    \lambda_e & = (N-1) \mu_{1P} \times \frac{\lambda_e}{ (N-1) \mu_{1P}}\approx (N-1) \mu_{1P} \rho, \\
    \lambda_s & = N \mu_{1P} \times \frac{\lambda_s}{ N \mu_{1P}}\approx N \mu_{1P} \rho. 
\end{align}
Consequently, the primary user's optimization problem is as stated below.

\begin{problem}[Primary device's MF optimization problem]\label{problem:primary_user_MF}
    \begin{align*}
        \min_{(p_P,\mu_{1P},\mu_{2P}) \in [0,1] \times \mathbb{R}^2} &  J_{P,\rho}(p_P,\mu_{1P},\mu_{2P}) \\
        &  \hspace{-16mm}\mbox{s.t.}~~ ~0 \leq \mu_{1P} \leq P_{max} \\
        & \hspace{-9mm}0 \leq \mu_{2P} \leq f_{P,max},
    \end{align*} 
    where $ \!J_{P,\rho}(p_P,\mu_{1P},\mu_{2P})  \!\!:=\!\!J_{P}^{(N)}\!(p_P,\mu_{1P},\mu_{2P},\lambda_s) \!\mid_{\lambda_s \!=\! N \mu_{1P} \rho}$ denotes the cost function of the primary user as a function of its decision variables and the mean-field generated by the ``minor'' secondary users.
\end{problem}

Next, for a generic secondary user of type $\phi$ with packet arrivals distributed as $\text{Poi}(\lambda_\phi)$, we split the incomings using a mean $p_\phi$ i.i.d. Bernoulli distributed r.v. and the service time of the local processer distributed as $\text{exp}(\mu_{2,\phi})$ with $\mu_{2,\phi} \leq f_{\phi,max}$. Thus, we can state the generic secondary device's local optimization problem as follows.

\begin{problem}[Secondary device MF optimization problem]\label{problem:sec_user_MF}
    \begin{align*}\\[-3em]
        \min_{(p_\phi,\mu_{2,\phi}) \in [0,1] \times \mathbb{R}} &  J_{S_\phi,\rho}(p_\phi,\mu_{1,\phi},\mu_{2,\phi},p_P,\mu_{1P},\mu_{2P}) \\
        & \hspace{-9mm}\mbox{s.t.}~~ ~  0\leq \mu_{2,\phi} \leq f_{\phi,max},
    \end{align*} 
    where $\!J_{S_\phi,\rho}(p_\phi,\mu_{1,\phi}, \mu_{2,\phi},p_P,\mu_{1P},\mu_{2P})\! :=\! J_{S_i}^{(N)}(\bm{p_{ps}},\bm{\mu_{2,ps}},$ $\bm{\mu_{1P}}) \mid_{\lambda_e = (N-1)\mu_{1P}\rho, \lambda_s = N\mu_{1P} \rho}$.
\end{problem}

Consequently, analogous to the equitable-access MEC case, the MFG is defined using three operators, namely, the optimality operators for the primary and secondary user of type $\phi$, and the consistency operators as follows:
\begin{enumerate}
    \item Optimality for the primary user: 
    \begin{align*}
        (\hat{p}_P,\hat{\mu}_{1P},\hat{\mu}_{2P}) & = \bar \Psi_1(\rho) := \text{argmin} ~J_{P,\rho}(p_P,\mu_{1P},\mu_{2P})
    \end{align*}
    subject to the constraints in Problem \ref{problem:primary_user_MF}.
    
    \item Optimality for the secondary user of type $\phi, \forall \phi \in \Phi$:
     \begin{align*}
    \!\! (\hat{p}_\phi,\hat{\mu}_{2,\phi}) \!\! =\!\! \bar \Psi_2(\rho)\!\!:=\!\!\text{argmin} \!J_{S_\phi,\rho}\!(p_\phi,\!\mu_{1,\phi},\!\mu_{2,\phi},\!\hat p_P,\!\hat \mu_{1P},\!\hat \mu_{2P}) 
    \end{align*}
    subject to the constraints in Problem \ref{problem:sec_user_MF}.
    
    \item Consistency: $\hat \rho  = \bar \Psi_3(\hat{p}_\phi,\hat{\mu}_{1,\phi},\hat{\mu}_{2,\phi}) := \mathbb{E}_{\mathbb P(\phi)}\Big[\frac{\lambda_\phi \hat{\Bar{p}}_\phi }{\mu_{1P}}\Big].$
\end{enumerate}

Briefly, the optimality operator $\bar \Psi_1(\cdot)$ outputs an optimal policy $(\hat p_P,\hat \mu_{1P},\hat \mu_{2P})$ for the primary user for a given MF $\rho$. Subsequently, the optimality operator $\bar \Psi_2(\cdot)$ outputs an optimal policy ($\hat p_\phi,\hat \mu_{2,\phi}$) for a secondary user of type $\phi$ for the same value of $\rho$ and the obtained values of $(\hat p_P,\hat \mu_{1P},\hat \mu_{2P})$ from $\bar \Psi_1(\cdot)$. Finally, the consistency operator $\bar \Psi_3(\cdot)$ generates a new $\rho$ by using the  optimal policies obtained above using $\bar \Psi_1$ and $\bar \Psi_2$, and signifies the consistent behavior of the population of the secondary users. The MFE which constitutes the tuple of equilibrium policies of the primary user, ($p_{P,\text{MFE}},\mu_{1P,\text{MFE}},\mu_{2P,\text{MFE}}$), the secondary users for all types, $(p_{\phi, \text{MFE}}, \mu_{1,\phi,\text{MFE}}, \mu_{2,\phi,\text{MFE}})_{\forall \phi}$, and the equilibrium mean-field ($\rho_{\text{MFE}}$), is given by the fixed point of the composite map of $\Psi_1$, $\Psi_2$, and $\Psi_3$. We provide Algorithm \ref{alg:MM_MFGs} to compute the MM-MFE.
Briefly, for a given value of $\rho$, we first solve Problem \ref{problem:primary_user_MF} to compute an optimal policy for the primary user (line 7 of Algorithm \ref{alg:MM_MFGs}). Then, given this policy, we solve Problem \ref{problem:sec_user_MF} for the secondary users (line 8 of Algorithm \ref{alg:MM_MFGs}). Finally, we update the mean-field using the consistency condition (line 9 of Algorithm \ref{alg:MM_MFGs}). The MFE is then given 
by the fixed point of the composition of the maps $\Psi_1, \Psi_2$, and $\Psi_3$.

\begin{algorithm}[t]
    \caption{Fixed-point iteration for computing an MM-MFE policy}
    \label{alg:MM_MFGs}
    \begin{algorithmic}[1]
    \STATE {\textbf{Input:} $V_\phi,V_P,\eta_\phi,\eta_P,\lambda_\phi,\lambda_P,\mu_3, ~\forall \phi$ \hfill\# MEC parameters}
    \STATE {\textbf{Input:} $\varepsilon_1, \varepsilon_2,\varepsilon_3$ \hfill \# tolerance parameters}
    \STATE {\textbf{Input:} $\gamma_1,\gamma_2,\gamma_3$ \hfill \# Iteration step sizes}
    \STATE {Initialize: $\hat{\rho}^{(0)}$, $\sigma_\phi^{(0)}:= (p_\phi^{(0)},\mu_{1,\phi}^{(0)},\mu_{2,\phi}^{(0)}), \forall \phi$}
    \STATE {Initialize: $\sigma_P^{(0)}:= (p_P^{(0)},\mu_{1P}^{(0)},\mu_{2P}^{(0)})$}
    \WHILE{$|\hat{\rho}^{(m)} - \hat{\rho}^{(m-1)}| < \varepsilon_1$}
    \STATE Compute the optimal policy for the primary user using lines 7-18 of Algorithm \ref{alg:MFE} and Problem \ref{problem:primary_user_MF}
    \STATE Given the primary user's policy, compute the optimal policy for secondary users using lines 6-19 of Algorithm \ref{alg:MFE} and Problem \ref{problem:sec_user_MF}.
    \STATE $\hat{\rho}^{(k)} \leftarrow (1-\gamma_1)\hat{\rho}^{(k-1)} + \gamma_1 \mathbb{E}_{\mathbb P(\phi)}\Big[\frac{\lambda_\phi^{(k)} (1-\hat{p}_\phi^{(k)}) }{\mu_{1P}^{(k)}}\Big]$
    \ENDWHILE
    \STATE \textbf{Output: } $\hat{\rho}^{(m)}, \sigma_P^{(m)}, \sigma_\phi^{(m)}, ~\forall \phi.$
    \end{algorithmic}
\end{algorithm}

Now, we have completely defined the MEC problem with priority access using MM-MFGs. In the next section, we will provide simulations to extensively evaluate the performance of the proposed approach for both the MEC with equitable access case and the MEC with priority access.

\section{Performance Evaluation}
In this section, we perform a rigorous performance evaluation of our proposed MFG approach to design optimal offloading policies for users. Since this is the first work employing the mean-field games formulation, a direct comparative study with literature may not be possible at this time; thus we perform an extensive validation and refer to historically observed trends, wherever applicable.

\subsection{The MEC System With Equitable Access}
In this subsection, we provide numerical analysis for the MEC system with equitable access. We consider a homogeneous population for the first 4 studies and heterogeneous for the 5th study, for ease in illustration of the main concepts. 

\begin{figure}[t]
    \centerline{\includegraphics[width=0.95\columnwidth]{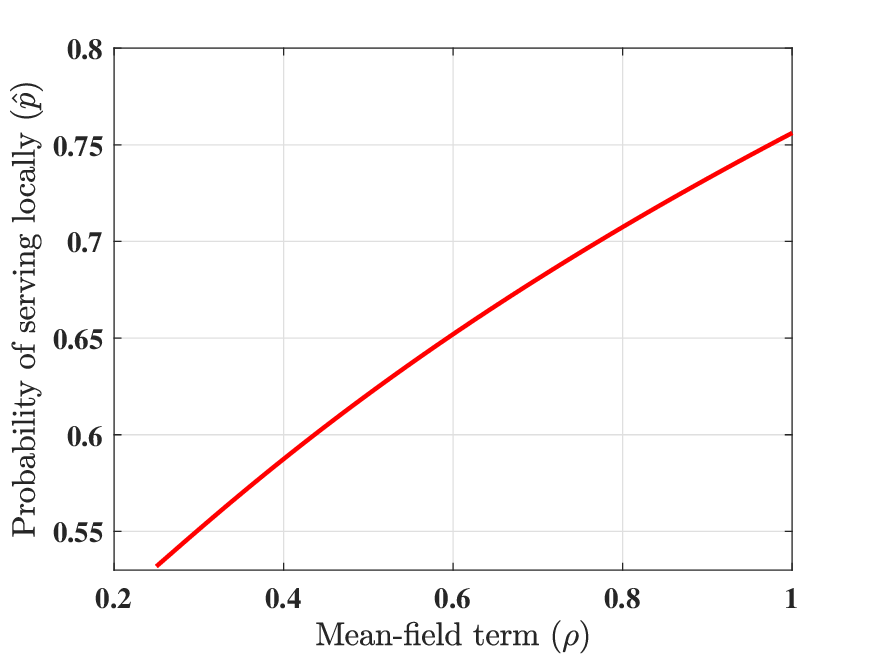}}
	\caption{\small{The optimal probability $\hat{p}$ as a function of the MF term $\rho$.}}
	\label{Fig:sim1}
\end{figure}

\textbf{1) Variation of $p$ vs $\rho$:} In the first numerical study of this subsection, we plot in Fig.~\ref{Fig:sim1}, the variation of local server usage versus the ES loading, denoted by the mean-field term $\rho$. We take the parameters  to be $V = 10,$ $ \eta = 0.5,$ $\lambda_\phi = 1,$ $ P_{\phi, max} = 1$, and $f_{\phi,max} = 0.8$. From Fig.~\ref{Fig:sim1}, we observe that as the mean load at the ES increases (on the $x$-axis), the optimal probability of using the local processor (on the $y$-axis) increases and that of offloading to the ES decreases. This should be expected since the devices care about the AoI $V \!=\!10$ times more than the average power consumed, as depicted by the value of $V$. Thus, if the ES is heavily loaded, the device is better-off serving tasks locally to incur a lower AoI. 

\begin{figure}[t]	
    \centerline{\includegraphics[width=0.99\columnwidth]{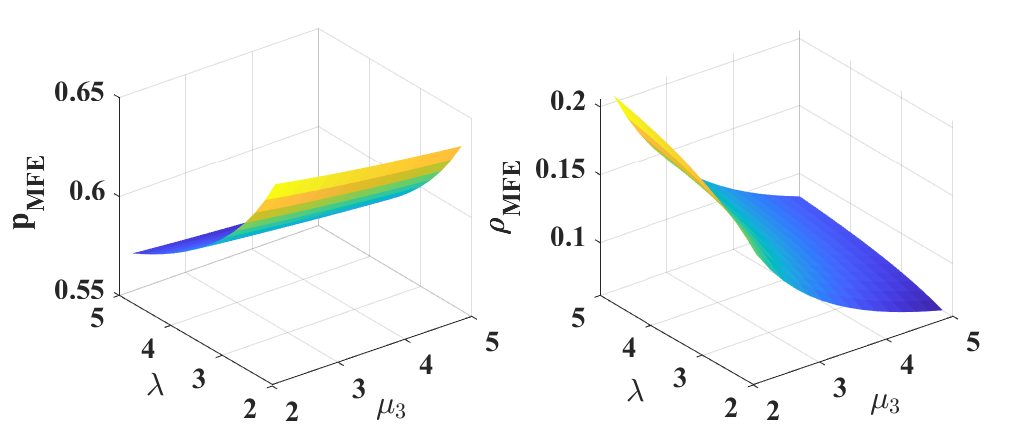}}
	\caption{\small{Variation of the equilibrium probability $p_{\text{MFE}}$ and equilibrium load $\rho_{\text{MFE}}$ with generic arrival rate $\lambda$ and service rate $\mu_3$.}}
	\label{Fig:sim2}
\end{figure}

\textbf{2) Effect of $\lambda$ and $\mu_3$ on MFE: }Next, in Fig.~\ref{Fig:sim2}, we plot the variation of MFE as a function of the arrival rate $\lambda$ and the service rate of the ES $\mu_3$. We consider the same set of parameters as in the previous study. In Fig.~\ref{Fig:sim2}, we observe that the equilibrium loading at the ES decreases (as seen from the right subplot) as $\mu_3$ increases with the decreasing probability of processing locally. In other words, as we increase $\mu_3$, the total number of packets processed at the edge server increases. However, since the service rate $\mu_3$ increases, the load on the edge server $\rho_{\text{MFE}}$ still decreases. The decrease in $\rho_{\text{MFE}}$ is low due to the fact that high $V$ restrains users from incurring too much staleness as a result of offloading. Furthermore, the equilibrium offloading probability $1-p_{\text{MFE}}$ increases with increasing arrival rate for given ES service rate $\mu_3$ to accommodate the higher arrival rate, which consequently also increases $\rho_{\text{MFE}}$. 

\begin{figure}[t]	
    \centerline{\includegraphics[width=0.9\columnwidth]{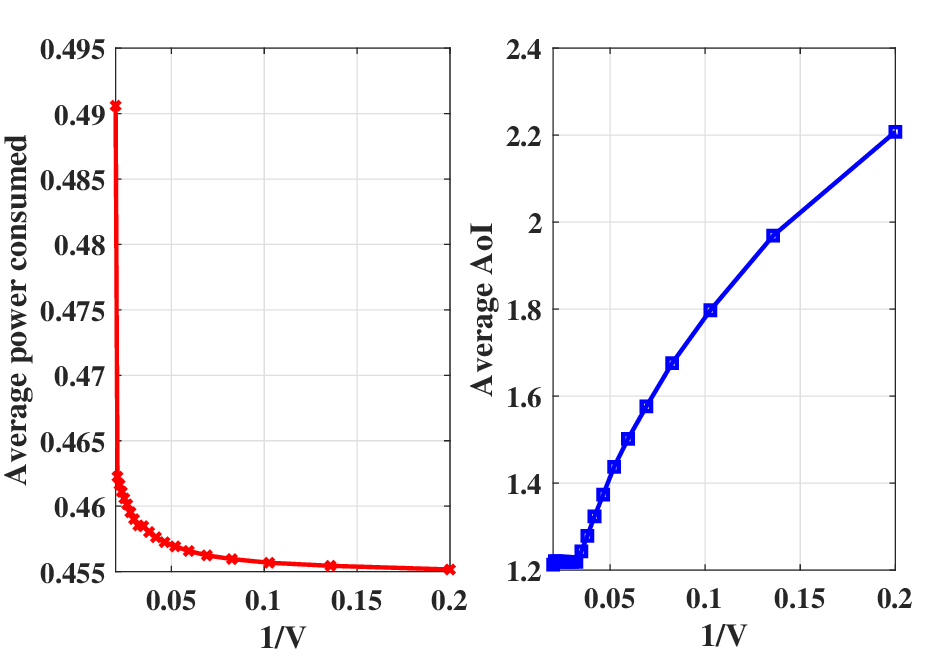}}
	\caption{\small{The average power and average AoI per user at equilibrium as a function of the inverse weighting parameter $1/V$ for $N=60$.}}
    \vspace{-2mm}
	\label{Fig:sim3}
\end{figure}

\textbf{3) Effect of $V$ on the power consumed and AoI incurred:} Next, in Fig. \ref{Fig:sim3}, we study the relationship between the average power consumption/average AoI incurred by the devices at equilibrium and the inverse weighting parameter $1/V$. We take the parameters to be $\eta = 0.02$, $\lambda = 10$, $\mu_3 = 15$, $P_{\phi,max} = 1$, $f_{\phi,max} = 0.8$, and $N=60$. We see from the left plot in Fig.~\ref{Fig:sim3} that the consumed power varies inversely with $1/V$ and converges to the value 0.455. Meanwhile, the average AoI incurred increases with $1/V$, as seen from the right plot. We would like to mention a couple of noteworthy points here. First, the trend observed through these plots closely resembles the $[O(V),O(1/V)]$ relationship observed in the works \cite{mao2016power,maharjan2019}. A direct comparison between our work and theirs, however, is not fair since the nature of optimization differs (static in our case and dynamic in theirs) and the solution approaches used are widely contrasting (distributed in our case and centralized in theirs). Nevertheless, it is worth noting that the resource allocation method used in their work causes an increase in the time complexity of computation with the number of users, and thus simulating a large number of agents can be challenging (as seen from their simulation results). In contrast, our MFG approach allows us to compute low complexity solutions, which can be scaled to a possibly large-user population. Additionally, during the same simulation, we also observed that after $V$ becomes sufficiently large ($V \geq 30$), the local processor at the device saturates to its upper operating frequency of $0.8$, and consequently, it requires an additional facility (other than the local one) to proceed with the computations, and prevent the AoI from increasing tremendously. This demonstrates the significance of an MEC system where the additional ES service keeps the AoI in check even when the local processor saturates. The same observation is also noted in detail for the next study.

\begin{figure}[t]	
    \centerline{\includegraphics[width=\columnwidth]{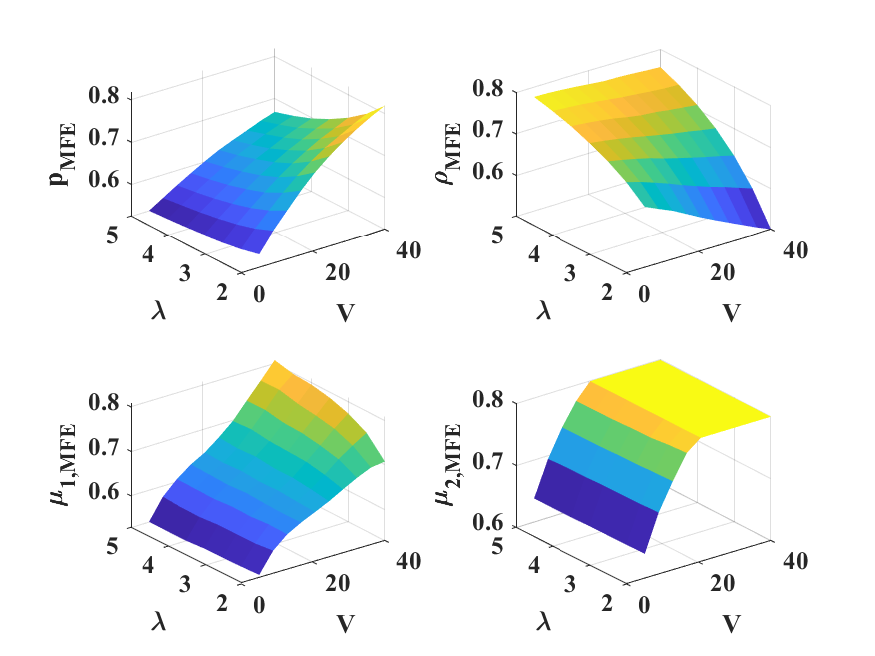}}
	\caption{\small{Variation of the equilibrium policy and equilibrium load $\rho_{\text{MFE}}$ with generic arrival rate $\lambda$ and weighting constant $V$.}}
	\label{Fig:sim4}
\end{figure}

\textbf{4) Effect of $\lambda$ and $V$ on MFE:} Next, in Fig.~\ref{Fig:sim4}, we study the dependence of the MFE solution on the arrival rates and the weighting constant for the same set of parameters as for the above study. We first observe that higher weighting on the AoI incentivizes the devices to process locally thereby increasing $p_{\text{MFE}}$. However,  $p_{\text{MFE}}$ increases only slightly with increasing $\lambda$. This is due to the fact that the local processor saturates at its maximum service rate after which more preemptions start causing an increase in the AoI, and hence it is suggested that we increase the transmissions to the ES as well. The latter phenomenon is apparent from the top right plot which shows that the ES loading increases for increasing $\lambda$ for a given $V$. Also, from the top right plot, we observe that for high arrival rates, even though $V$ increases, the ES equilibrium load increases. This suggests that once the local processor saturates, there is no option left for the device but to offload to the ES to manage the high rate incoming tasks. Further, we can also observe the local processor saturation from the bottom right plot with increase in $V$ and $\lambda$ at its maximum operating frequency $\mu_{1,\text{MFE}}$ at equilibrium. This then causes a simultaneous increase in the transmission of packets, which can be seen by an increase in the transmission power $\mu_{1,\text{MFE}}$ of the transmitter, to compensate for the increase in average AoI due to saturation. This observation also helps us demonstrate the necessity of an MEC system, and verify the ``quality of experience'' of computation in an MEC system since without an ES, the AoI would keep increasing due to local saturation thereby leading to a computation bottleneck at the device.

\textbf{5) Comparison of NE and MFE policies:} Finally, we compare the performance of the proposed MFE approach (computed using Algorithm \ref{alg:MFE}) with the corresponding Nash equilibrium (NE) performance (computed using Algorithm \ref{alg:Nash_game}). The users are distributed within 5 types, each with different values of $\lambda_\phi, f_{\phi,max}$ and $P_{\phi,max}$ and with $V = 10$ and $\eta = 0.5$ as in the earlier studies. The results of the values obtained are tabulated in Table \ref{table:NE_vs_MFE} for three different values of $N=30,40$ and $50$. From the same, we observe that the MFE policy approximates the NE policy reasonably well in the sense that the both of them yield nearly the same performance for each user.

\begin{table}[t]
    \begin{center}
    \begin{tabular}{|c|c|c|c|}
    \hline
    $N$ & {$(\mu_{1,\text{NE}},\mu_{2,\text{NE}},p_\text{NE})$}         
    & {($\mu_{1,\text{MFE}},\mu_{2,\text{MFE}},p_{\text{MFE}}$)} \\ \hline
    30 & (0.5603,0.6007,0.5202) & (0.5647,0.6474,0.5701) \\ \hline
    40 & (0.5644,0.6117,0.5233) & (0.5452,0.6196,0.5681) \\ \hline
    50 & (0.5719,0.6363,0.5314)  & (0.5322,0.5969,0.5655) \\ \hline
    \end{tabular}
    \end{center}
    \label{table:NE_vs_MFE}

    \begin{center}
    \begin{tabular}{|c|c|c|c|}
    \hline
    $N$ & {cost under NE policy}         
    & {cost under MFE policy} \\ \hline
    30 & 16.0739 & 13.5142 \\ \hline
    40 & 21.5955 & 18.9599 \\ \hline
    50 & 26.4713  & 24.6980 \\ \hline
    \end{tabular}
    \end{center}
    \caption{\small{The first table shows the comparison of a local NE policy with the local MFE policy obtained for different values of $N$ while the second table shows the comparison of the cost per user under the two policies.}}
    \label{table:NE_vs_MFE}
\end{table}

Our numerical study of the equitable access MEC game is now complete. Next, we provide numerical analysis for the MEC system comprised of primary and secondary users and will numerically compute similar equilibrium policies.

\subsection{The MEC with Priority Access}
\begin{figure}[t]
    \centerline{\includegraphics[width=0.9\columnwidth]{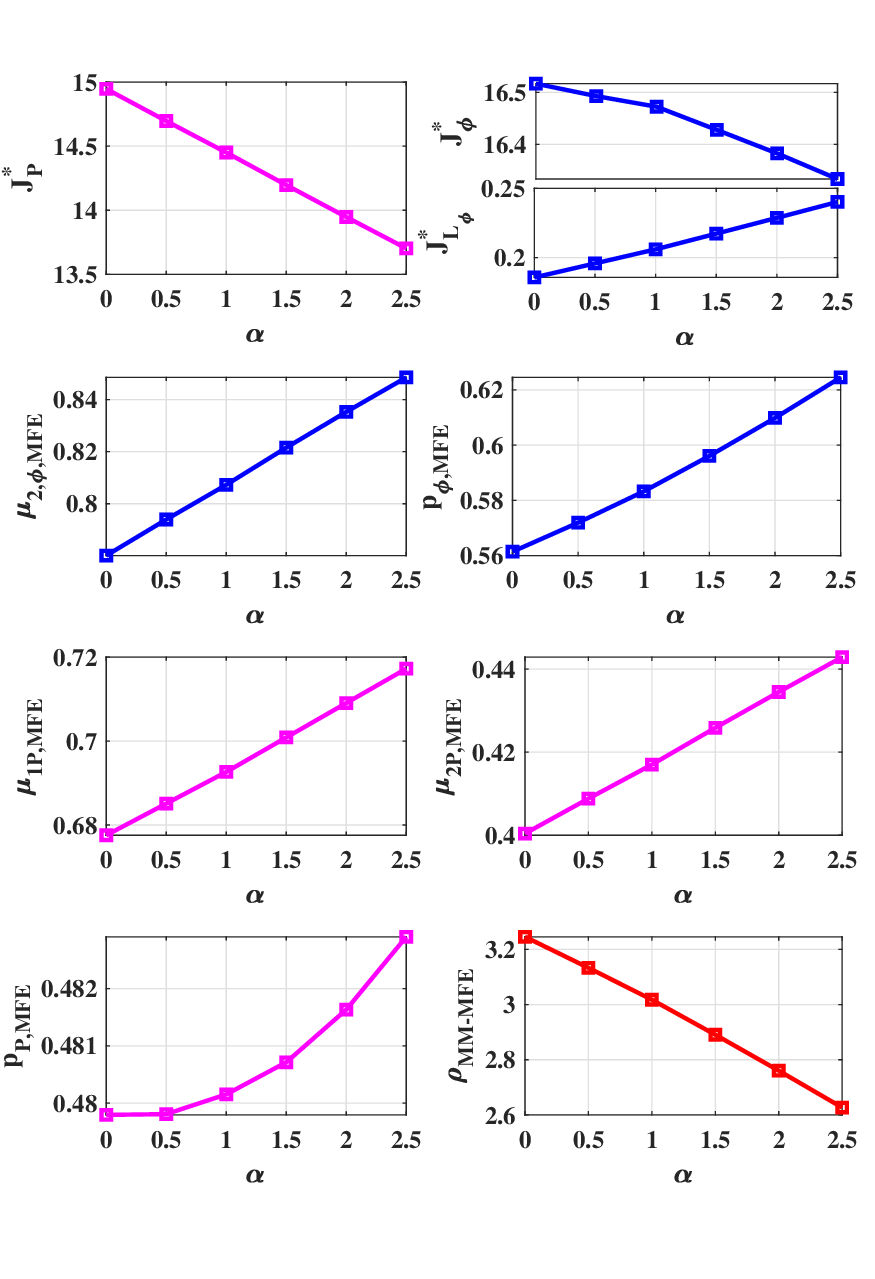}}
        \vspace{-0.7cm}
	\caption{\small{Variation of the equilibrium policies, the equilibrium costs and the equilibrium load for the primary and the secondary users with the price parameter $\alpha$.}}
	\label{Fig:sim_alpha}
    \vspace{-0.4cm}
\end{figure}

\textbf{1) Effect of $\alpha$ on  MFE and associated costs:} In the first study of this subsection, we plot Fig. \ref{Fig:sim_alpha} which shows the effect of variations of the pricing parameter $\alpha$ on different quantities in the MM-MEC MFG. We take $N=30$, $f_{P,max} = 3$, $f_{\phi,max} = 1.5$, $V=10$, $\eta = 0.5$, $P_{max} = 10$, $\lambda_P = 4$, $\lambda_\phi = 5$, and $\mu_3 = 15$. The initial conditions for the primary user were set to $(\mu_{1P}, \mu_{2P}, p_P) = (0.6,0.3,0.5)$ and those for the secondary users to be $(\mu_{2,\phi}, p_\phi) = (0.6,0.5)$. The quantities relating to the primary user are plotted in pink color, the ones related to the secondary user are plotted in blue and the mean load is plotted in red color. First, we observe that, with increasing price per unit, the equilibrium cost $J^*_P$ of the primary user decreases due to increasing revenue earned. Meanwhile, the local processing cost for the secondary user $J^*_{L_\phi}$ increases since with a higher price charged by the primary user, the secondary user prefers to use its local processor more. This is further confirmed by its equilibrium policy constituting its local processing frequency $\mu_{2,\phi,\text{MFE}}$  and its probability of serving locally $p_{\phi,\text{MFE}}$, both of which also increase. The decrease in the total cost $J_\phi^*$ of the secondary user is due to a better average AoI as a consequence of fewer offloads, which counteracts the increase in the price charged by the primary user. The equilibrium policy $(\mu_{1P,\text{MFE}}, \mu_{2P,\text{MFE}}, p_{P,\text{MFE}})$ shows only a slight change for different values of $\alpha$. This is due to the change in the equilibrium load $\rho_{\text{MFE}}$ which affects the final equilibrium values. The value of $\rho_{\text{MFE}}$ shows a decrease due to the lesser and lesser affinity of the secondary users toward offloading, thereby causing a decrease in the equilibrium load at the primary user's transmitter. 

\begin{figure}[t]	
    \centerline{\includegraphics[width=\columnwidth]{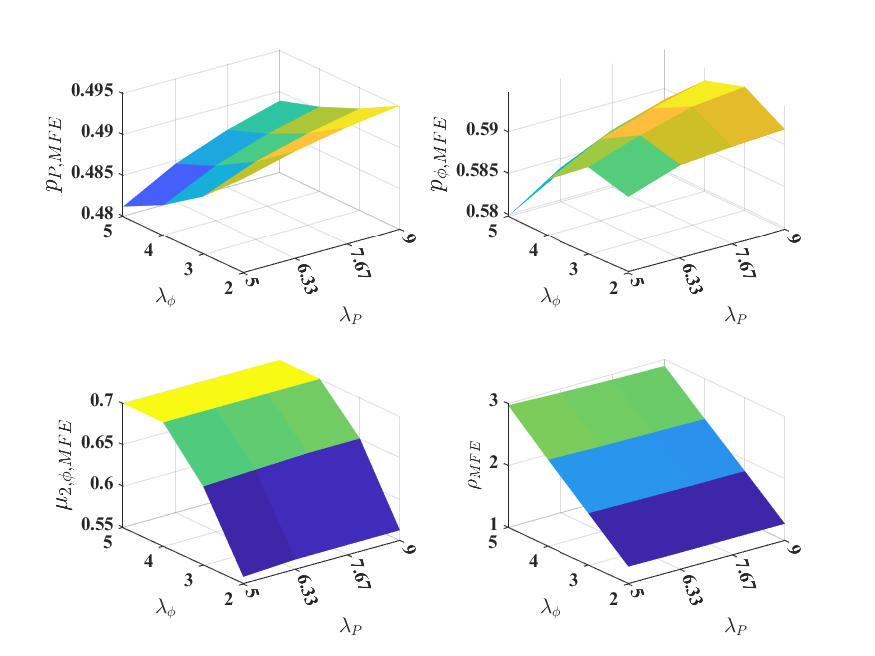}}
	\caption{\small{Variation of the equilibrium policies of the primary and secondary users arrival rates $\lambda_P$ and $\lambda_\phi$.}}
	\label{Fig:sim_MM_MFE}
\end{figure}

\textbf{2) Effect of $\lambda_P$ and $\lambda_\phi$ on MFE:} Next, in Fig. \ref{Fig:sim_MM_MFE}, we study the effect of variations in primary and secondary users' incoming arrival rates $\lambda_P,$ $\lambda_\phi$, respectively, on the MFE policies. We take $V=10,$ $\eta=0.5$, $\alpha = 1$, $\mu_3 = 15$, $P_{max} = 2$, $f_{\phi,max} = 0.7$, $f_{P,max} = 0.5$, and $N=30$. Furthermore, the initial conditions were set to $(\mu_{1P}, \mu_{2P}, p_P) = (0.6,0.3,0.5)$ and $(\mu_{2,\phi}, p_\phi) = (0.2,0.5)$. First, we observe from the bottom left plot that as the arrival rates increase, the secondary user's local processor operating frequency at equilibrium, $\mu_{2,\phi,\text{MFE}}$, increases to accommodate the increasing incoming tasks. Further, the probability $p_{\phi,\text{MFE}}$ of serving locally shows a nominal decrease (from the top right plot) to compensate for increasing preemptions at the local processor due to increased arrival rate $\lambda_\phi$. The primary user's probability of serving locally also shows a nominal increase to accommodate the increasing arrivals caused by the increasing average AoI cost as a result of increased preemptions at the transmitter of the primary user. Finally, the mean-loading at equilibrium, $\rho_{\text{MFE}}$, increases with increasing arrival rate of the secondary users while being almost unaffected by the increasing arrival rate of the primary user (as the mean effect is only generated by the population of secondary users). 

\textbf{3) Effective resource utilization:} In our final experiment, we study the effective resource utilization with and without the presence of secondary users.  We take $V=10,$ $\eta=0.5,$ $N=30,$ $\mu_3 = 15,$ $f_{P,max} = 0.5,$ $P_{max}= 2$, and $\lambda_P = 2$. First, we note that for the case where only the primary user used its transmitter $T_P$, and hence, the ES, the effective busy period of $T_P$ was obtained as $t_{T_p,1} = 0.6248$ by solving for the device's optimization problem (as defined in Problem \ref{problem:primary_user}) with $\lambda_s=0$ and $\alpha = 0$. Thus, $T_P$ was idling $\sim 38\%$ of the time. However, for the case, when there are secondary users (with $\alpha=1,$ $f_{\phi,max} = 0.7$, and $\lambda_\phi=1$) which can utilize the ES through $T_P$, the effective busy period of $T_P$ was calculated as $t_{T_p} = t_{T_p,1} + t_{T_p,2} = 0.6032 + 0.3864 = 0.9896$, where $t_{T_P,1},$ $t_{T_P,2}$ were defined above Problem \ref{problem:primary_user}. Thus, the idling time is only $\sim 1\%$ in this case. The effective utilization of the shared transmitter has thus increased by a significant $37\%$, leading to a better utilization of the ES, which is in line with our motivation behind the primary-secondary based-MEC system with priority access to begin with, which was inspired by the CRN technology. 

\section{Discussion \& Conclusion}
In this work, we have considered two different MEC architectures with and without priority access, comprised of $N$ devices where each device is assisted by an edge server to service computation intensive tasks. To alleviate the issues of scalability under a large population of users, we have provided low complexity algorithms for both the architectures to compute fully distributed approximate Nash equilibrium policies for each device by setting up the optimization problem of each device toward balancing the power consumed as a result of local processor usage and task offloading, and the timeliness of information affected by the usage of a shared edge computing facility in the presence of a large number of other users. In the process, we invoked techniques from stochastic hybrid systems (SHS) theory to obtain necessary equations to characterize the average AoI of each user, and consequently employed the mean-field game (and major-minor mean-field game) paradigms to allow for distributed decision making at each user by making use of local information and the statistical information about the system. Finally, we performed extensive numerical evaluations, leading to insights into how system parameter variations affect the equilibrium offloading policies for the users, the average AoI incurred, and the power consumed by the devices.  

Our results pave a way for exciting future research directions. The first one that deserves mention is to relax the discipline of LCFS-P to ones with the presence of (a possibly finite capacity) buffer facility for the users. This would then allow for a packet in service to be processed and delivered while other packets still wait in the queue. Considering such a model, however, makes the AoI analysis more complex due to the increase in state space of the overall MEC system with the length of the buffer. It would be interesting to examine how this impacts the users' equilibrium offloading policies as we explore new trade-offs introduced by the buffer length as a parameter.

Another important direction would be to consider a networked MEC architecture where each user can reach the ES through neighbouring devices' transmitters over multiple hops. This would then lead to interesting notions of device-to-device/peer-to-peer communication within a joint competitive-cooperative game framework, whereby each device wants to balance its selfish objectives whilst also requiring to cooperate with other users by allowing them to utilize their transmitters for establishing communication with the ES. Such a problem formulation may help address the challenges of how to price external device utilization, how to discourage free riding, and how to obtain Nash equilibrium policies for each user within a large user setting. One of the promising frameworks for this setting is that of graphon mean-field games \cite{caines2019graphon,caines2021graphon}, which are counterparts of \textit{standard} mean-field games employed in this work. The former allows for heterogeneous interactions between agents occurring over a network which can be modeled by using techniques from graph theory and consequently attempts to compute equilibrium decision policies for each user in the game.

\bibliographystyle{IEEEtran}
\bibliography{references}

\begin{thebibliography}{10}
\providecommand{\url}[1]{#1}
\csname url@samestyle\endcsname
\providecommand{\newblock}{\relax}
\providecommand{\bibinfo}[2]{#2}
\providecommand{\BIBentrySTDinterwordspacing}{\spaceskip=0pt\relax}
\providecommand{\BIBentryALTinterwordstretchfactor}{4}
\providecommand{\BIBentryALTinterwordspacing}{\spaceskip=\fontdimen2\font plus
\BIBentryALTinterwordstretchfactor\fontdimen3\font minus
  \fontdimen4\font\relax}
\providecommand{\BIBforeignlanguage}[2]{{%
\expandafter\ifx\csname l@#1\endcsname\relax
\typeout{** WARNING: IEEEtran.bst: No hyphenation pattern has been}%
\typeout{** loaded for the language `#1'. Using the pattern for}%
\typeout{** the default language instead.}%
\else
\language=\csname l@#1\endcsname
\fi
#2}}
\providecommand{\BIBdecl}{\relax}
\BIBdecl

\bibitem{aggarwal2024mean}
S.~Aggarwal, M.~A. uz~Zaman, M.~Bastopcu, S.~Ulukus, and T.~Ba{\c{s}}ar,
  ``Fully decentralized task offloading in multi-access edge computing
  systems,,'' in \emph{2024 IEEE Global Communications Conference Workshop -
  6GComm (Globecom 2024 Workshop - 6GComm)}, Cape Town, South Africa, 8-12
  December 2024.

\bibitem{mach2017mobile}
P.~Mach and Z.~Becvar, ``Mobile edge computing: A survey on architecture and
  computation offloading,'' \emph{IEEE Communications Surveys \& Tutorials},
  vol.~19, no.~3, pp. 1628--1656, March 2017.

\bibitem{hua2018energy}
M.~Hua, Y.~Huang, Y.~Wang, Q.~Wu, H.~Dai, and L.~Yang, ``Energy optimization
  for cellular-connected multi-{UAV} mobile edge computing systems with
  multi-access schemes,'' \emph{Journal of Communications and Information
  Networks}, vol.~3, no.~4, pp. 33--44, December 2018.

\bibitem{muhammad2021minimizing}
A.~Muhammad, I.~Sorkhoh, M.~Samir, D.~Ebrahimi, and C.~Assi, ``Minimizing age
  of information in multiaccess-edge-computing-assisted {IoT} networks,''
  \emph{IEEE Int. of Thgs. Jrnl.}, vol.~9, no.~15, pp. 13\,052--13\,066, Dec.
  2021.

\bibitem{kaul2012real}
S.~Kaul, R.~Yates, and M.~Gruteser, ``Real-time status: How often should one
  update?'' in \emph{IEEE Infocom}, March 2012.

\bibitem{huang2004uplink}
M.~Huang, P.~E. Caines, and R.~P. Malham{\'e}, ``Uplink power adjustment in
  wireless communication systems: A stochastic control analysis,'' \emph{IEEE
  Trans. on Autom. Control}, vol.~49, no.~10, pp. 1693--1708, October 2004.

\bibitem{huang2006large}
M.~Huang, R.~P. Malham{\'e}, and P.~E. Caines, ``Large population stochastic
  dynamic games: closed-loop {M}ckean-{V}lasov systems and the {N}ash
  {C}ertainty {E}quivalence principle,'' \emph{Communications in Information \&
  Systems}, vol.~6, no.~3, pp. 221--252, 2006.

\bibitem{huang2007large}
M.~Huang, P.~E. Caines, and R.~P. Malham{\'e}, ``Large-population cost-coupled
  {LQG} problems with nonuniform agents: individual-mass behavior and
  decentralized $\varepsilon $--{N}ash equilibria,'' \emph{IEEE Transactions on
  Automatic Control}, vol.~52, no.~9, pp. 1560--1571, September 2007.

\bibitem{lasry2007mean}
J.~M. Lasry and P.~L. Lions, ``Mean field games,'' \emph{Japanese Journal of
  Mathematics}, vol.~2, no.~1, pp. 229--260, March 2007.

\bibitem{wang2014mean}
Y.~Wang, F.~R. Yu, H.~Tang, and M.~Huang, ``A mean field game theoretic
  approach for security enhancements in mobile ad hoc networks,'' \emph{IEEE
  Trans. on Wireless Communications}, vol.~13, no.~3, pp. 1616--1627, January
  2014.

\bibitem{aggarwal2023weighted}
S.~Aggarwal, M.~A.~U. Zaman, M.~Bastopcu, and T.~Ba{\c{s}}ar, ``Weighted age of
  information based scheduling for large population games on networks,''
  \emph{IEEE Journal on Selected Areas in Information Theory}, vol.~4, pp.
  682--697, November 2023.

\bibitem{aggarwal2023large}
S.~Aggarwal, M.~A. uz~Zaman, M.~Bastopcu, and T.~Ba{\c{s}}ar, ``Large
  population games on constrained unreliable networks,'' in \emph{2023 62nd
  IEEE Conf. on Decision and Control (CDC)}, 2023, pp. 3480--3485.

\bibitem{priyadarshi2024techniques}
R.~Priyadarshi, R.~R. Kumar, and Z.~Ying, ``Techniques employed in distributed
  cognitive radio networks: a survey on routing intelligence,''
  \emph{Multimedia Tools and Applications}, pp. 1--52, 2024.

\bibitem{nourian2013}
M.~Nourian and P.~E. Caines, ``$\epsilon$--{N}ash mean field game theory for
  nonlinear stochastic dynamical systems with major and minor agents,''
  \emph{SIAM Jrnl. on Control and Optim.}, vol.~51, no.~4, pp. 3302--3331,
  2013.

\bibitem{firoozi2017execution}
D.~Firoozi and P.~E. Caines, ``The execution problem in finance with major and
  minor traders: A mean field game formulation,'' \emph{Advances in Dynamic and
  Mean Field Games: Theory, Applications, and Numerical Methods}, pp. 107--130,
  2017.

\bibitem{mao2016power}
Y.~Mao, J.~Zhang, S.~Song, and K.~B. Letaief, ``Power-delay tradeoff in
  multi-user mobile-edge computing systems,'' in \emph{2016 IEEE Global
  Communications Conference (GLOBECOM)}.\hskip 1em plus 0.5em minus 0.4em\relax
  IEEE, 2016, pp. 1--6.

\bibitem{wang2017joint}
F.~Wang, J.~Xu, X.~Wang, and S.~Cui, ``Joint offloading and computing
  optimization in wireless powered mobile-edge computing systems,'' \emph{IEEE
  Trans. on Wireless Communications}, vol.~17, no.~3, pp. 1784--1797, December
  2017.

\bibitem{maharjan2019}
S.~Mao, S.~Leng, S.~Maharjan, and Y.~Zhang, ``Energy efficiency and delay
  tradeoff for wireless powered mobile-edge computing systems with multi-access
  schemes,'' \emph{IEEE Transactions on Wireless Communications}, vol.~19,
  no.~3, pp. 1855--1867, 2020.

\bibitem{jia2022lyapunov}
Y.~Jia, C.~Zhang, Y.~Huang, and W.~Zhang, ``Lyapunov optimization based mobile
  edge computing for internet of vehicles systems,'' \emph{IEEE Transactions on
  Communications}, vol.~70, no.~11, pp. 7418--7433, September 2022.

\bibitem{neely2022stochastic}
M.~Neely, \emph{Stochastic Network Optimization with Application to
  Communication and Queueing Systems}.\hskip 1em plus 0.5em minus 0.4em\relax
  Springer Nature, May 2022.

\bibitem{kuang2019age}
Q.~Kuang, J.~Gong, X.~Chen, and X.~Ma, ``Age-of-information for
  computation-intensive messages in mobile edge computing,'' in \emph{IEEE
  WCSP}, October 2019.

\bibitem{liu2021optimizing}
L.~Liu, X.~Qin, X.~Xu, H.~Li, F.~R. Yu, and P.~Zhang, ``Optimizing information
  freshness in {MEC}-assisted status update systems with heterogeneous energy
  harvesting devices,'' \emph{IEEE Internet of Things Journal}, vol.~8, no.~23,
  pp. 17\,057--17\,070, April 2021.

\bibitem{lin2019computation}
L.~Lin, X.~Liao, H.~Jin, and P.~Li, ``Computation offloading toward edge
  computing,'' \emph{Proc. of the IEEE}, vol. 107, no.~8, pp. 1584--1607, 2019.

\bibitem{feng2022computation}
C.~Feng, P.~Han, X.~Zhang, B.~Yang, Y.~Liu, and L.~Guo, ``Computation
  offloading in mobile edge computing networks: A survey,'' \emph{Journal of
  Network and Computer Applications}, vol. 202, p. 103366, 2022.

\bibitem{zhou2020partial}
S.~Zhou and W.~Jadoon, ``The partial computation offloading strategy based on
  game theory for multi-user in mobile edge computing environment,''
  \emph{Computer Networks}, vol. 178, p. 107334, 2020.

\bibitem{ning2020mobile}
Z.~Ning, P.~Dong, X.~Wang, X.~Hu, L.~Guo, B.~Hu, Y.~Guo, T.~Qiu, and R.~Y.~K.
  Kwok, ``Mobile edge computing enabled 5{G} health monitoring for internet of
  medical things: A decentralized game theoretic approach,'' \emph{IEEE Journal
  on Sel. Areas in Comm.}, vol.~39, no.~2, pp. 463--478, 2020.

\bibitem{pham2022partial}
X.-Q. Pham, T.~Huynh-The, E.-N. Huh, and D.-S. Kim, ``Partial computation
  offloading in parked vehicle-assisted multi-access edge computing: A
  game-theoretic approach,'' \emph{IEEE Transactions on Vehicular Technology},
  vol.~71, no.~9, pp. 10\,220--10\,225, 2022.

\bibitem{teng2022game}
H.~Teng, Z.~Li, K.~Cao, S.~Long, S.~Guo, and A.~Liu, ``Game theoretical task
  offloading for profit maximization in mobile edge computing,'' \emph{IEEE
  Transactions on Mobile Computing}, vol.~22, no.~9, pp. 5313--5329, 2022.

\bibitem{zhou2022stackelberg}
H.~Zhou, Z.~Wang, N.~Cheng, D.~Zeng, and P.~Fan, ``Stackelberg-game-based
  computation offloading method in cloud--edge computing networks,'' \emph{IEEE
  Internet of Things Journal}, vol.~9, no.~17, pp. 16\,510--16\,520, 2022.

\bibitem{Kang2024}
Y.~Kang, Y.~Zhu, D.~Wang, Z.~Han, and T.~Başar, ``Joint server selection and
  handover design for satellite-based federated learning using mean-field
  evolutionary approach,'' \emph{IEEE Transactions on Network Science and
  Engineering}, vol.~11, no.~2, pp. 1655--1667, March-April 2024.

\bibitem{olmez2022modeling}
S.~Y. Olmez, S.~Aggarwal, J.~W. Kim, E.~Miehling, T.~Ba{\c{s}}ar, M.~West, and
  P.~G. Mehta, ``Modeling presymptomatic spread in epidemics via mean-field
  games,'' in \emph{2022 American Control Conference (ACC)}.\hskip 1em plus
  0.5em minus 0.4em\relax IEEE, 2022, pp. 3648--3655.

\bibitem{carmona2020applications}
R.~Carmona, ``Applications of mean field games in financial engineering and
  economic theory,'' \emph{arXiv preprint arXiv:2012.05237}, 2020.

\bibitem{al2015joint}
A.~Y. Al-Zahrani, F.~R. Yu, and M.~Huang, ``A joint cross-layer and colayer
  interference management scheme in hyperdense heterogeneous networks using
  mean-field game theory,'' \emph{IEEE Transactions on Vehicular Technology},
  vol.~65, no.~3, pp. 1522--1535, 2015.

\bibitem{banez2020mean}
R.~A. Banez, H.~Tembine, L.~Li, C.~Yang, L.~Song, Z.~Han, and H.~V. Poor,
  ``Mean-field-type game-based computation offloading in multi-access edge
  computing networks,'' \emph{IEEE Transactions on Wireless Communications},
  vol.~19, no.~12, pp. 8366--8381, 2020.

\bibitem{yates2018status}
R.~D. Yates, ``Status updates through networks of parallel servers,'' in
  \emph{2018 IEEE Intl. Symp. on Info. Theory (ISIT)}, 2018, pp. 2281--2285.

\bibitem{kaul2020timely}
S.~K. Kaul and R.~D. Yates, ``Timely updates by multiple sources: The {M}/{M}/1
  queue revisited,'' in \emph{2020 54th Annual Conference on Information
  Sciences and Systems (CISS)}.\hskip 1em plus 0.5em minus 0.4em\relax IEEE,
  2020, pp. 1--6.

\bibitem{gai2016packet}
Y.~Gai, H.~Liu, and B.~Krishnamachari, ``A packet dropping mechanism for
  efficient operation of {{M}/{M}/1} queues with selfish users,''
  \emph{Computer Networks}, vol.~98, pp. 1--13, 2016.

\bibitem{Bastopcu_distortion}
M.~Bastopcu and S.~Ulukus, ``Age of information for updates with distortion:
  Constant and age-dependent distortion constraints,'' \emph{IEEE/ACM
  Transactions on Networking}, vol.~29, no.~6, pp. 2425--2438, 2021.

\bibitem{Buyukates_dist_comp}
B.~Buyukates and S.~Ulukus, ``Timely distributed computation with stragglers,''
  \emph{IEEE Trans. on Comm.}, vol.~68, no.~9, pp. 5273--5282, 2020.

\bibitem{Li_MEC_UAV}
H.~Li, J.~Zhang, H.~Zhao, Y.~Ni, J.~Xiong, and J.~Wei, ``Joint optimization on
  trajectory, computation and communication resources in information freshness
  sensitive mec system,'' \emph{IEEE Transactions on Vehicular Technology},
  vol.~73, no.~3, pp. 4162--4177, 2024.

\bibitem{Song_MEC}
X.~Song, X.~Qin, Y.~Tao, B.~Liu, and P.~Zhang, ``Age based task scheduling and
  computation offloading in mobile-edge computing systems,'' in \emph{2019 IEEE
  Wireless Communications and Networking Conference Workshop (WCNCW)}, 2019,
  pp. 1--6.

\bibitem{sathyavageeswaran2024timely}
N.~Sathyavageeswaran, R.~D. Yates, A.~D. Sarwate, and N.~Mandayam, ``Timely
  offloading in mobile edge cloud systems,'' \emph{arXiv preprint
  arXiv:2405.07274}, 2024.

\bibitem{yates2020age}
R.~D. Yates, Y.~Sun, D.~R. Brown, S.~K. Kaul, E.~Modiano, and S.~Ulukus, ``Age
  of information: An introduction and survey,'' \emph{IEEE Jrnl. on Sel. Areas
  in Comm.}, vol.~39, no.~5, pp. 1183--1210, May 2021.

\bibitem{sun2022age}
Y.~Sun, I.~Kadota, R.~Talak, and E.~Modiano, \emph{Age of Information: A New
  Metric for Information Freshness}.\hskip 1em plus 0.5em minus 0.4em\relax
  Springer Nature, 2022.

\bibitem{boyd2004convex}
S.~P. Boyd and L.~Vandenberghe, \emph{Convex Optimization}.\hskip 1em plus
  0.5em minus 0.4em\relax Cambridge University Press, 2004.

\bibitem{hespanha2006modelling}
J.~P. Hespanha, ``Modelling and analysis of stochastic hybrid systems,''
  \emph{IEE Proc.-Control Theory and Apps.}, vol. 153, no.~5, pp. 520--535,
  September 2006.

\bibitem{yates2018age}
R.~D. Yates and S.~K. Kaul, ``The age of information: Real-time status updating
  by multiple sources,'' \emph{IEEE Transactions on Information Theory},
  vol.~65, no.~3, pp. 1807--1827, September 2018.

\bibitem{norris1998markov}
J.~R. Norris, \emph{Markov Chains}.\hskip 1em plus 0.5em minus 0.4em\relax
  Cambridge University Press, 1998, no.~2.

\bibitem{bacsar1998dynamic}
T.~Ba{\c{s}}ar and G.~J. Olsder, \emph{Dynamic Noncooperative Game
  Theory}.\hskip 1em plus 0.5em minus 0.4em\relax SIAM, 1998.

\bibitem{wright2015coordinate}
S.~J. Wright, ``Coordinate descent algorithms,'' \emph{Mathematical
  Programming}, vol. 151, no.~1, pp. 3--34, 2015.

\bibitem{caines2019graphon}
P.~E. Caines and M.~Huang, ``Graphon mean field games and the {GMFG} equations:
  $\varepsilon$-{N}ash equilibria,'' in \emph{2019 IEEE 58th Conference on
  Decision and Control (CDC)}.\hskip 1em plus 0.5em minus 0.4em\relax IEEE,
  2019, pp. 286--292.

\bibitem{caines2021graphon}
------, ``Graphon mean field games and their equations,'' \emph{SIAM Journal on
  Control and Optimization}, vol.~59, no.~6, pp. 4373--4399, 2021.

\end{thebibliography}
\end{document}